\documentclass[prb,twocolumn,amsmath,amssymb,floatfix,showpacs,superscriptaddress,showkeys]{revtex4}

\usepackage{bm}

\usepackage{amssymb,amsmath}
\usepackage{graphics}
\usepackage{subfigure}
\usepackage{epstopdf}
\usepackage{epsfig}
\usepackage{color}
\usepackage{amsfonts}
\usepackage{bm}
\usepackage{amscd}
\usepackage{epsfig}
\usepackage{subfigure}
\usepackage{color}
\usepackage{color}
\definecolor{darkgray}{rgb}{0.3,0.3,0.3}
\definecolor{gray}{rgb}{0.5,0.5,0.5}
\definecolor{yellow}{rgb}{.4,.4,0}
\definecolor{orange}{rgb}{1,0.5,0}
\definecolor{darkgreen}{rgb}{0,0.5,0}
\definecolor{darkblue}{rgb}{0,0,1}
\definecolor{darkred}{rgb}{0.5,0,0}
\definecolor{purple}{rgb}{0.35,0,0.35}

\def\sig{{\mbox{\boldmath{$\sigma$}}}}

\begin{document}
\title{Spin-polarized dynamic transport in   tubular two-dimensional electron gases }

\author{E. A. Rothstein}

\email{rotshtei@post.bgu.ac.il}

\affiliation{Department of Physics, Ben Gurion University, Beer
Sheva 84105, Israel}

\author{B. Horovitz}

\affiliation{Department of Physics, Ben Gurion University, Beer
Sheva 84105, Israel}

\author{ O. Entin-Wohlman}

\affiliation{Department of Physics, Ben Gurion University, Beer
Sheva 84105, Israel}

\affiliation{Raymond and Beverly Sackler School of Physics and Astronomy, Tel Aviv University, Tel Aviv 69978, Israel}


\author{A. Aharony}


\affiliation{Department of Physics, Ben Gurion University, Beer
Sheva 84105, Israel}

\affiliation{Raymond and Beverly Sackler School of Physics and Astronomy, Tel Aviv University, Tel Aviv 69978, Israel}


\date{\today}

\begin{abstract}

The ac conductance of a finite tubular two-dimensional electron gas  is studied in the presence of the Rashba spin-orbit interaction.  When the tube is coupled to two reservoirs, that interaction splits the  steps in the dc current, introducing energy ranges with spin-polarized currents. For this setup, we calculate the current-current correlations (the noise spectrum) and show that the existence of these dc spin-polarized currents can be deduced from the shot noise. We also find that the Wigner-Smith time delay is almost unaffected by the spin-orbit interaction. When the tube is coupled to a single  reservoir, we calculate the quantum capacitance and the charge-relaxation resistance, and find that they exhibit singularities near the openings of new channels.


\end{abstract}


\maketitle

\section{Introduction}

\label{Intro}

Quite generally, the wavy nature of electrons together with the ensuing interference effects
determine a large variety of quantum-coherence phenomena in quantum wires and dots.
The electronic spin, being weakly coupled to other degrees of freedom in bulk materials,  becomes an active player in these small systems. \cite{Winkler}
This  is due to the enhanced spin-orbit interaction induced by the Rashba effect, \cite{Rashba} that can be  also modified experimentally. \cite{NITTA1,ENGELS} In multiply-connected mesoscopic systems, the   effect of the  spin-orbit interaction resembles that of an orbital magnetic field, \cite{Meir} paving the way to possible intriguing interference-induced outcomes. \cite{Aharony} Indeed, there is an on-going vast experimental effort 
to study spin-orbit coupling effects in tubular systems, notably carbon nanotubes \cite{Kuemmeth,Feuillet} but  also DNA   and other long chiral  molecules.  \cite{Xie} However, the spin-orbit interaction
in carbon nanotubes may not be predominantly of the Rashba type. \cite{Ando} It  appears that core-shell semiconducting devices are more suitable to explore interference effects resulting from the Rashba spin-orbit coupling.

Core-shell nanowires  comprise a thin layer (shell) surrounding a core in a tubular geometry. \cite{core} While the charge carriers in these devices can be confined either to the core \cite{Tilburg} or to the shell, \cite{Rieger} it is clear that interference effects are more pronounced in   the second case. \cite{Jung}
Here we focus on this configuration, disregarding the core altogether.
The electrical conductance of a finite-length tubular two-dimensional electron gas (2DEG)  subject to the Rashba spin-orbit interaction has been analyzed 
exploiting scattering theory  in the context of the Landauer formula.
\cite{ENTIN,Rosdahl} In this paper we study the frequency-dependent conductance of such a system, when it is  connected to a tubular lead (or leads) where the electrons move ballistically.  Measurements and calculations of the dynamic conductance supplement  those of  dc transport properties: while the latter yield the transmission of the mesoscopic system, the former  contain in addition information related to  the phases of the scattering matrix.
The reason being that the ac quantities are given in terms of elements of the scattering matrix, and not only by their absolute values.

In the first part of the paper,  Sec. \ref{rc}, 
we derive the dynamic conductance ${\cal G}(\omega)$  of a gated tubular 2DEG
connected to a single electronic reservoir by a tubular lead (see Fig. \ref{figrc}) and study its low-frequency properties. This quantity, often referred to as admittance, 
is customarily presented in the form \cite{BUTTIKER3}
\begin{align}
{\cal G}(\omega )\simeq-i\omega{\cal C}+\omega^{2}{\cal C}^{2}{\cal R}\ ,
\label{gexp}
\end{align}
where $\omega$ is the frequency of the driving field.
The expansion (\ref{gexp}) introduces the ``quantum capacitance" ${\cal C}$ and the charge-relaxation resistance ${\cal R}$, both being topics of active research (see Sec. \ref{rc}). We present a detailed calculation of those for a tubular 2DEG, and in particular relate the capacitance to the Friedel phase and the charge accumulated in the tube.
In the second part of the paper, Sec. \ref{SPEC}, the tubular 2DEG is connected   to two reservoirs by two ballistic tubular leads, see Fig. \ref{figfull}. We calculate the various current-correlation functions,   the shot noise and the Wigner-Smith time-delay matrix. 
In both Sec. \ref{rc} and Sec. \ref{SPEC} we first present the analytic expressions and then exemplify the results by several plots. 
The paper is supplemented by three appendices: the first details the derivation of the reflection matrix for the setup depicted in  Fig. \ref{figrc}, the second discusses the limit where the scattering system is large enough for the frequency to exceed the level spacing, and the third gives the details of the calculation of the scattering matrix for the configuration shown in Fig. \ref{figfull}.

Our calculations are based on the scattering-matrix approach \cite{BUTTIKER} for noninteracting electrons.   Electron-electron interactions are not taken into account;
however,  much of the physics in the Coulomb-blockade regime is believed to be captured
by such models, with an effective Hartree-like energy \cite{Hackenbroich}  (which can be incorporated into the our calculation quite easily).  This simplification allows us to carry out the analysis analytically. Another major simplification stems from the geometry of the core-shell systems: in a perfect tube, the linear Rashba spin-orbit interaction does not mix the transverse channels.  For this reason, the effective magnetic field due to the spin-orbit coupling is in a sense analogous to that of an orbital magnetic field, similarly leading to interference phenomena.

Our research was motivated by the quest to detect hallmarks of the spin-orbit coupling in the ac properties of a mesoscopic conductor.  The results in Sec. \ref{rc} 
show that  the spin cannot be considered as another transport channel since the spin-orbit interaction mingles the two spin directions and 
causes a dependence of the quantization direction
on the scattering energy. We also find there that 
the universal value of the charge-relaxation resistance predicted in Ref. \onlinecite{BUTTIKER3} arises only when the transport occurs via the lowest-energy channel and  is lost when higher-energy channels are included; this property, however,  does not necessitate the spin-orbit coupling (though the latter does modify the results, see the discussion in Sec. \ref{rc}). In a way,  the conclusions drawn in Sec. \ref{SPEC} are much more rewarding; in particular they indicate possibilities to induce and detect    spin-polarized currents.  The reason is related to the effect of the Rashba spin-orbit interaction on the transmission. As a function of the energy of the scattered electron, one of the spin channels may be blocked, and then the transmitted current is polarized. \cite{ENTIN} This is reflected in the dc conductance, and also in the shot noise (see Sec. \ref{SPEC}).  We also find that this polarization can be manipulated by, e.g., a gate  voltage. Thus, tubular core-shell systems appear to be  interesting candidates for spintronic devices.

\section{Quantum capacitance and charge-relaxation resistance}

\label{rc}

Ever since the experimental verification  \cite{GABELLI}
of the prediction made in Ref. 
\onlinecite{BUTTIKER3} (see also Refs. \onlinecite{BUTTIKER2} and \onlinecite{PRETRE})
concerning the universal value of the charge-relaxation resistance, there has been considerable interest in the low-frequency electrical properties of mesoscopic conductors.
Here we examine those for a tubular mesoscopic conductor, in which spin-orbit interaction of the Rashba type is effective.

\begin{figure}[htp]
\includegraphics[width=7.5cm]{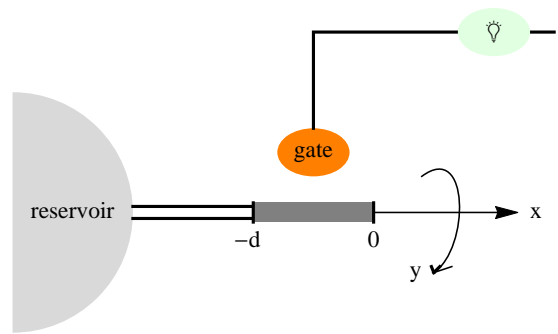}
\caption{(Color online) Tubular mesoscopic RC circuit. An ac source excites a periodic accumulation of charges on the gate, and the latter affects the charges on the mesoscopic cylinder (the dark region confined by the vertical thick lines) and thus creates an ac current  flowing in the cylindrical  lead (the light region of the tube) connecting the conductor to an electronic reservoir.  A spin-orbit interaction of the Rashba type is operative in the mesoscopic cylinder, in the region $-d\leq x\leq 0$.} 
\label{figrc}
\end{figure}

The system we study is depicted in Fig. \ref{figrc}: a mesoscopic cylinder,   placed along the $\hat{\bf x}$ direction in the region $-d\leq x\leq 0$,   is separated by a delta-function potential  from the region $x\leq -d$ where the spin-orbit interaction vanishes and the electrons move ballistically. This potential is characterized by a single parameter $\zeta$     (measured in momentum units; units in which $\hbar =1$ are used). When $\zeta$ is very large, the interface approaches  the tunnel-junction limit.  We assume that on the right  (at $x\geq 0$) the cylinder is totally pinched off.  Thus the system is described by the Hamiltonian 
\begin{align}
\label{HRC}&{\cal H}=\frac{1}{2m}(p^{2}_{x}+p^{2}_{y}
)+\frac{\zeta}{m}\delta (x+d)\\
&+\Big (\Theta(-x)\Theta(x+d)\frac{\alpha}{2m}(p^{}_{y}\sigma^{}_{x}-p^{}_{x}\sigma^{}_{y})+{\rm Hc}\Big )\ , \nonumber
\end{align}
where ${\bf p}=(p_{x},p_{y})$ is the two-dimensional momentum operator, $\alpha$
is the strength of the spin-orbit interaction (in momentum units) and $\sigma_{x,y}$ are the Pauli matrices.
The coordinate system is specified in Fig. \ref{figrc}. Note that  the hermitian conjugation in the last term of Eq. (\ref{HRC}) yields a delta-function term, $-(\alpha/2m)\sigma_{y}\delta (x+d)$. This term is  crucial for ensuring current continuity across the interface at $x=-d$.

The ac conductance of noninteracting electrons moving through a mesoscopic conductor can be presented in terms of the scattering matrix of the latter.  \cite{BUTTIKER3,PRETRE} For the setup displayed in Fig. \ref{figrc}, there is only a reflection matrix; its derivation is given in Appendix \ref{App1}. In this geometry,    the wave vector along $\hat{\bf y}$,  denoted $q$,   is fixed by  the periodic boundary  condition of this direction
\begin{align}
q=\frac{2\pi n}{L}\ ,\ {\rm with}\ \ n=0,\pm1,\pm 2,\ldots\ , 
\label{q}
\end{align}
where $L$ is the circumference of the tube.
The wave vector along the propagation direction $\hat{\bf x}$ is  fixed by the energy $E$ of the scattered electron. 
In the ballistic region it is
\begin{align}
k=\sqrt{2mE-q^{2}}\ . 
\label{k}
\end{align}
In the mesoscopic conductor
where the spin-orbit interaction is effective it is expedient to measure the energy from $\alpha^{2}/(2m)$ (by adding this constant to the Hamiltonian of the scatterer). One finds that the longitudinal wave vector 
takes two values, 
\begin{align}
k^{}_{\rm u,d}=\sqrt{(\sqrt{2mE}\pm\alpha)^{2}-q^{2}}\ .
\label{kud}
\end{align}
[The $+$ ($-$) sign belongs to  $k_{\rm u}$ ($k_{\rm d}$).]
Note that while $k_{\rm u}$ is always real (since $q^{2}\leq 2m E$), 
$k_{\rm d}$ may be purely imaginary
and then one of the waves in the region $-d \leq x\leq 0$ 
is evanescent. \cite{EAV,ENTIN}

In our  tubular geometry the
transverse channels are not coupled, and the scattering matrix splits into blocks each describing the scattering matrix for a certain value of $q$, i.e., for a certain  $n$. Due to the presence of the spin-orbit coupling   spin is not conserved and therefore those blocks are 4$\times$4 matrices. The reflection ${\bf R}$ is hence a 2$\times$2 matrix. We show in Appendix \ref{App1} that 
\begin{align}
{\bf R}(q)=
[-1+{\bf F}^{-1}(q)]\ ,
\label{SM}
\end{align}
where the matrix ${\bf F}$ is too cumbersome to be reproduced here. It is  shown in Appendix \ref{App1}  that ${\bf F}$ can be decomposed into
\begin{align}
{\bf F}(q)=F^{}_{1}(q)+\sigma^{}_{x}F^{}_{x}(q)+\sigma^{}_{z}F^{}_{z}(q)\ ,
\label{FF}
\end{align}
where  the explicit expressions for the components are given in  Eqs. (\ref{FC}) and (\ref{OM}). In particular, for $q=0$ $F_{x}$ and $F_{z}$ vanish, and ${\bf R}_{q=0}$ is proportional to the unit matrix, i.e., the spin effects disappear.  Indeed, when the energy of the scattered electron is too low to support a nonzero transverse mode, the motion becomes effectively one-dimensional and then  the spin-orbit interaction can be eliminated by a gauge transformation,  \cite{ENTIN}
${\cal U}{\cal H}{\cal U}^{\dagger}$, with ${\cal U}=\exp [-i\alpha\sigma _{y}x]$. This cancels the $\alpha-$term in the Hamiltonian;
the second boundary condition in  Eqs. (\ref{bcd})
acquires then the term $\partial_{x}{\cal U}$ which cancels the $i\alpha\sigma_{y}$ term there.

The ac conductance  in the linear-response regime, ${\cal G}(\omega )$, is given by
\begin{align}
{\cal G}(\omega )=&
\frac{e^{2}}{2\pi}\int^{\infty}_{-\infty}dE\frac{f(E)-f(E+\omega)}{\omega}\nonumber\\
&\times\sum_{q}{\rm Tr}\{(1-{\bf R}^{\dagger}_{q}(E){\bf R}^{}_{q}(E+\omega)\}\ , 
\label{acg}
\end{align}
where the trace is carried out in spin space. (It is written  here in terms of the reflection alone; the more general form is given in Sec. 
\ref{QF}.) The Fermi function,  $f(E)=[\exp [(E-\mu )/(k^{}_{\rm B}T)]+1]^{-1}$, describes the distribution of the electrons in the reservoir, with $\mu$ being the chemical potential there; below we confine our discussion to zero temperature and therefore $\mu$ is equal to the Fermi energy,  $\mu=E_{F}$.
The universal value of the charge-relaxation resistance  discovered in Ref. \onlinecite{BUTTIKER3}
emerges upon comparing the low-frequency expansion of Eq. (\ref{acg}) [given in Eq. (\ref{gexp})]
with the ac conductance,
\begin{align}
{\cal G}_{a}^{}(\omega )=-i\omega {\cal C}+\omega^{2}{\cal C}^{2}{\cal R}+{\cal O}(\omega^{3})\ ,
\label{aca}
\end{align}
of a conventional capacitor whose capacitance is equal to 
${\cal C}$ and which is connected in series to a  resistor whose dc  resistance is ${\cal R}$.
For a single-channel scatterer, the reflection  is just a phase factor,  ${\bf R}(E)=\exp[i\phi (E)]$. One then  finds for ${\cal R}$ the value $\pi/(2e^{2})$,  half of the quantum unit of the resistance; it is independent of  the scattering properties of the conductor. 
The capacitance, on the other hand, is given by ${\cal C}=(e^{2}/2\pi)\phi '(E_{F})$, where 
$\phi '$ is the energy derivative of the 
reflection phase at the Fermi energy.

The case of the tubular conductor is different from the one treated in  Ref. \onlinecite{BUTTIKER3}: first there are numerous transverse channels, and second there are the spin effects, rendering ${\bf R}(q)$ 
a unitary matrix (instead of being just a phase factor).
Nonetheless, the capacitance  can still be expressed in terms of phases. Indeed, 
the eigenvalues of  ${\bf R}(q)$  are $\exp[i\phi^{}_{1,2}]$
(for brevity we omit in some of the expressions the explicit dependence on $q$)
\begin{align}
e^{i\phi^{}_{1,2}}=-
\Big (1-\frac{1}{\lambda^{}_{1,2}}\Big )\ ,
\label{ep}
\end{align}
where $\lambda_{1,2}$ are the eigenvalues of ${\bf F}$, Eq. (\ref{FF}), 
\begin{align}
\lambda^{}_{1.2}=F^{}_{1}\pm\sqrt{F^{2}_{1}-{\rm det}({\bf F})}\ .
\label{lam}
\end{align}
In particular we note that for the lowest transverse channel $q=0$, the two eigenvalues are identical, $\phi_{1}(q=0)=\phi_{2}(q=0)$ (see the discussion in 
Appendix \ref{App1}). The
quantum capacitance 
of the tubular 2DEG is  given by
\begin{align}
{\cal C}=\frac{e^{2}}{2\pi}\sum_{q}\Big (\frac{\partial \phi^{}_{1}(E,q)}{\partial E}+
\frac{\partial \phi^{}_{2}(E,q)}{\partial E}\Big )\Big |^{}_{E=E^{}_{F}} \ .
\label{c}
\end{align}
On the other hand, the charge-relaxation resistance, 
\begin{align}
{\cal R}=\frac{e^{2}}{4\pi{\cal C}^{2}}
\sum_{q}{\rm Tr}\Big (\frac{d{\bf R}^{\dagger}}{dE}\frac{d{\bf R}}{dE}\Big )\ ,
\label{r}
\end{align}
involves also the energy-derivatives of the quantization axis. 
This can be seen by presenting  the  reflection in the form
\begin{align}
{\bf R}=e^{i\phi}
e^{-i\theta\hat{\bf n}\cdot\sig}\ ,
\label{SSM}
\end{align}
where $\sig$ is the vector of the Pauli matrices and the angles $\phi_{1, 2}$, Eq. (\ref{ep}), are given by
$\phi\pm\theta$. 
We show in Appendix \ref{App1} [see Eq. (\ref{RAP})]
that 
\begin{align}
\exp[i\phi]=\sqrt{\frac{{\rm det}({\bf F}^{\ast}_{})}{{\rm det}({\bf F})}}\ ,
\end{align}
the unit vector $\hat{\bf n}$ (that depends on the energy), around which the spin rotates in spin space because of the spin-orbit coupling is the direction of the vector $(F_{x},0,F_{z})$, 
and
\begin{align}
\cos\theta=\frac{|F^{}_{1}|^{2}+F^{2}_{x}+F^{2}_{z}}{\sqrt{(|F^{}_{1}|^{2}+F^{2}_{x}+F^{2}_{z})^{2}-F^{2}_{x}-F^{2}_{z}}}\ .
\end{align}
Exploiting the form Eq. (\ref{SSM}) of the reflection matrix, one finds that
\begin{align}
{\rm Tr}\Big (\frac{d{\bf R}^{\dagger}}{dE}\frac{d{\bf R}}{dE}\Big )
&=
\Big (\frac{\partial \phi_{1}}{\partial E}\Big )^{2}+
\Big (\frac{\partial \phi_{2}}{\partial E}\Big )^{2}\nonumber\\
&
+2\frac{\partial\hat{\bf n}}{\partial E}\cdot\frac{\partial\hat{\bf n}}{\partial E}
\sin^{2}\frac{\phi^{}_{1}-\phi^{}_{2}}{2}\Big |^{}_{E=E^{}_{F}} \ .
\label{sec}
\end{align}
The appearance of the last term in Eq. (\ref{sec}) is a direct result of the interference of the two spin directions. 
Obviously, the universal value of the charge-relaxation resistance that is independent of the details of the scatterer is obtained when the Fermi energy is so low that only the lowest transverse channel is excited. Then there remains only the $q=0$ term in the sum, for which $\phi_{1}=\phi_{2}$.

The capacitance ${\cal C}$ may be related to the number, $N_{\rm D}$, of displaced electrons around the scatterer (at energy $E$).  \cite{BUTTIKER3}
According to  the relation derived by Langer and Ambegaokar, \cite{LA} the Friedel sum-rule
\begin{align}
N^{}_{D}(E)=\frac{1}{2i\pi}{\rm Tr}\ln [
{\bf R}(E^{}_{})]\ ,
\label{sumr}
\end{align}
gives  $N_{\rm D}$ in terms of  the {\em full} reflection  (see also Ref. \onlinecite{Taniguchi}).
By exploiting the identity $\ln {\rm det}(\mathbf{M})={\rm Tr}\ln \mathbf{M}$ where $\mathbf{M}$ is an arbitrary matrix,   we find
\begin{align}
N^{}_{D}(E)&=\frac{1}{2 i\pi}\ln \prod_{q}e^{i[
{\phi}_{1}(E,q)+
{\phi}_{2}(E,q)]}\nonumber\\
&=\frac{1}{2\pi}\sum_{q}[
{\phi}_{1}(E,q)+
{\phi}_{2}(E,q)]\ .
\label{nd}
\end{align}
Comparing 
Eq. (\ref{nd}) with Eq. (\ref{c}) shows that 
\begin{align}
{\cal C}=e^{2}\frac{dN^{}_{D}}{dE}\ ,
\label{DOS}
\end{align}
in agreement with Ref. \onlinecite{BUTTIKER3} (see the discussion at the end of Appendix \ref{App1}). 
However, $N_{D}(E)$,  as well as its energy derivative, 
 are meaningful only when the scattering phases are measured outside the system, typically asymptotically,  \cite{LA,Taniguchi} whereas we  measure our phases relative to x=-d; 
this  definition may cause the capacitance to attain negative  values
(see Appendix A).

\begin{figure}

 \includegraphics[width=7.5cm]{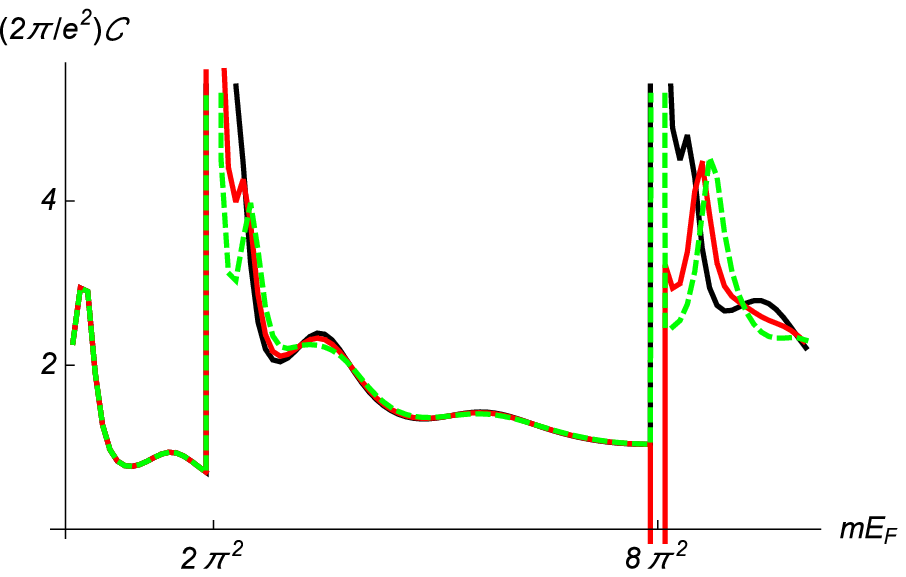} 
  \includegraphics[width=7.5cm]{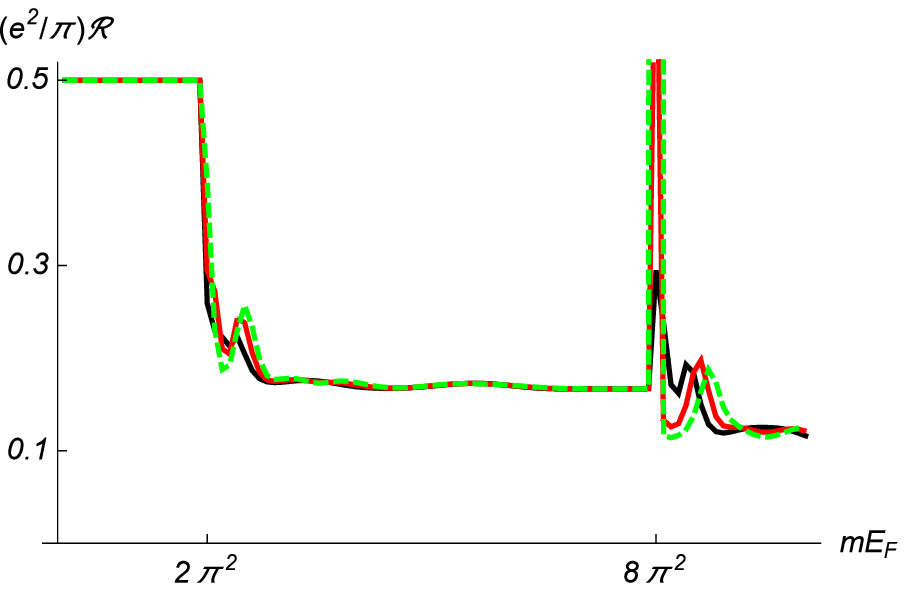} 
  \caption{(Color online)  Upper  panel: the quantum capacitance, Eq. (\ref{c}); lower panel: the charge-relaxation resistance, Eq. (\ref{r}),   as a function of the Fermi energy (normalized by $m$, in 
units of $L^{-2}$) and  $\zeta=0.6$ (in units of $L^{-1}$). The solid (black) curve is for $\alpha=  0.1$, the smaller-dashed (red) one is for  $\alpha =0.2$, and the large-dashed (green) curve is for $\alpha = 0.3$ (in units of $L^{-1}$).
In all our figures $d=L$.}  \label{FIG AZ}
\end{figure}

The 
quantum  capacitance  and the  charge-relaxation resistance are plotted  in Figs. 
\ref{FIG AZ} as a function of the Fermi energy.
The 
striking aspect of these figures  are the sharp extrema in both the quantum capacitance and the charge-relaxation resistance, albeit the rather low barrier between the scatterer and the lead ($\zeta$=0.6 in the figure).
The ones in the capacitance (see the upper panel in Fig. \ref{FIG AZ}) 
correspond to the standing waves in the tube (when  detached).  
An intriguing  point is the negativeness of ${\cal C}$ near the second step for $\alpha =0.2$ (in units of $L^{-1}$). The appearance of negative values depends on the  choice of parameters and also on the point along the $\hat{\bf x}-$axis 
relative to which the phase is measured.   
The latter feature implies that some charge has been displaced to the range $x<-d$.
Also note that the vanishing of ${\cal C}$ leads to        
formally a diverging ${\cal R}$, yet the measurable quantity [as well as the expansion parameter in Eq. (\ref{gexp})] is ${\cal C}^{2}{\cal R}$, which is finite when ${\cal C}=0$. The charge-relaxation resistance  itself  is a useful object when it is quantized (below the first step) or nearly quantized (not too close to other steps).


The resistance for $n=0$ (i.e., for $q=0$) is constant, reflecting the universal value of the charge-relaxation resistance discussed above.  For finite values of $q$ [i.e., $n\neq 0$, Eq. (\ref{q})]  the ``base line" of the charge-relaxation 
resistance is approximately at ${\cal R}=\pi/[2(2n+1)e^{2}]$, as if comprising $2n +1$ resistors in parallel,   of magnitude $\pi/(2e^{2})$ each,   in agreement    with B\"{u}ttiker {\it et al.} \cite{BUTTIKER2}
(recall the degeneracy of $q$). 
The spin-orbit coupling 
removes the degeneracy of the longitudinal wave function, and causes (when strong enough, see the large-dashed curve in Fig. \ref{FIG AZ}) the splitting of the second peak of ${\cal R}$.

We note in passing that the measured capacitance is different from  ${\cal C}$ as defined in Eq. (\ref{c}).  \cite{BUTTIKER3, BUTTIKER10} There the capacitance is deduced from  the current response $dI$ at the lead (at $-\infty$) to the voltage $dV$ relative to the potential on the probed region $dU$, i.e. $dI={\cal G}(\omega)(dV-dU)=-i\omega {\cal C}(dV-dU)+{\cal O}(\omega^2)$. On the other hand, the measured capacitance ${\cal C}_{m}$ is defined by $dI=-i\omega {\cal C}_{m}(dV-dV')$ where $dV'$ is the potential change on a gate near the probed region. The potential change $dV'$ generates locally a change in the charge such that $dI=-i\omega {\cal C}_e(dU-dV')$ where ${\cal C}_e$ is known as the geometric capacitance. Eliminating $dU$ one finds that $1/C_{m}=1/C_e+1/C$. \cite{BUTTIKER3, BUTTIKER10}

The low-frequency expansion leading to Eq. (\ref{c}) for the capacitance and Eq. (\ref{r}) for the charge-relaxation resistance has to be handled cautiously when the scattering tube is long enough  for the frequency to exceed the level spacing,  
$md^{2}\gg \omega^{-1}$. The reason is that then
the reflection matrix elements [as a function of $E$, $E+\omega$, see Eq. (\ref{acg})] are wildly oscillating.  
We examine this case in Appendix \ref{App1a} (ignoring the spin-orbit interaction for simplicity); in particular we show that the charge-relaxation resistance of the lowest transverse mode is $2\pi/(2e^{2})$, in agreement with Refs. \onlinecite{Mora}.

\section{The noise spectrum }
\label{SPEC}

\subsection{General expressions}
\label{genfull}

\begin{figure}[htp]
\includegraphics[width=7.5cm]{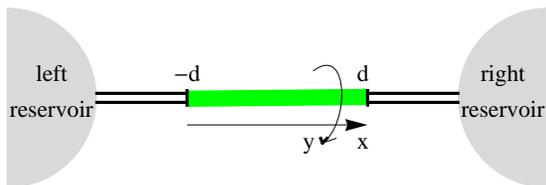}
\caption{(Color online) A tubular mesoscopic conductor [the dark (green) part of the cylinder] is connected to two reservoirs via leads (the light parts of the tube) where the electrons move ballistically. The region where spin-orbit interaction is active ($|x|<d$) is separated from the leads by two potential barriers (thick vertical lines). A magnetic field along  $\hat{\bf x}$  adds to the versatility of the device.} 
\label{figfull}
\end{figure}

Here we study the current-correlation functions, i.e., the noise spectrum of a tubular 2DEG subject to the Rashba interaction, see   Fig. \ref{figfull}. The tube   is placed along the $\hat{\bf x}-$direction and we include in the analysis the orbital effect of a magnetic field along  $\hat{\bf x}$, which might add versatility to the device.
The magnetic field  is specified by a flux $\Phi$ penetrating the cylinder. The Rashba-affected tubular 2DEG, confined to the region  $|x|\leq d$,  is separated from the cylindrical  leads by two tunnel junctions characterized by $\zeta_{L}$ and $\zeta_{R}$ for the left and the right barrier, respectively (in units of momentum). These leads are coupled each to an electronic reservoir, where the electronic distribution is
\begin{align}
f^{}_{\gamma}(E)=[e^{\beta (E-\mu_{\gamma})}+1]^{-1}\ ,\ \ \ \gamma =L \ \ {\rm or}\ \ R\ ,
\end{align}
with $\mu_{\gamma}$ being the chemical potential in reservoir $\gamma$. We assume that the reservoirs are not spin polarized, and therefore the Fermi functions do not depend on the spin index.

The Hamiltonian describing this setup is
\begin{align}
{\cal H}&=\frac{1}{2m} (p^{2}_{x}+p^{2}_{y})+\frac{\zeta^{}_{L}}{m}\delta (x+d)+\frac{\zeta^{}_{R}}{m}\delta (x-d)
\nonumber\\
&+\frac{\alpha}{2m}\Big (\Theta(d-|x|)(p^{}_{y}\sigma^{}_{x}-p^{}_{x}\sigma^{}_{y})+{\rm Hc}\Big )\ .\label{HAMFUL}
\end{align}
Once again we measure the energy $E$ in the region $ |x|\leq d$ with respect to $\alpha^{2}/(2m)$ [see comment after Eq. (\ref{k})].
The presence of a magnetic field along  $\hat{\bf x}$  modifies the $y$ component of the momentum, $p_{y}\rightarrow p_{y}-A_{y}$, where $A_{y}=e\Phi/(cL)$ is the vector potential in units of inverse length. As a result, the wave vector $q$ along $\hat{\bf y}$ given in Eq. (\ref{q}) is modified as well, 
$q\rightarrow q-\varphi$, where $\varphi\equiv A_{y}$. Below, we keep the notation $q$ for the transversal momentum, bearing in mind the shift caused  by the  magnetic flux.
The  current-current correlations (i.e., the noise spectrum) are expressed in terms of the scattering matrix of the scatterer, i.e., the $|x|\leq d$ region. \cite{BUTTIKER}
This matrix, pertaining to the Hamiltonian  (\ref{HAMFUL}), is derived in Appendix \ref{App2}.  


Within the scattering formalism, the time-dependent operator of the current leaving  lead $\chi $,  $\hat{I}_{\chi}(t)$ is \cite{BUTTIKER} ($\chi =L$ or $R$)
\begin{align}
\label{IT}
&\hat{I}^{}_{\chi}(t)=\frac{e}{2\pi}\sum_{\tilde{\sigma}}\int_{-\infty}^{\infty}dE
\int_{-\infty}^{\infty}dE'
e^{i(E-E')t}
\\
&\times\sum_{\gamma,\gamma '}\sum_{\sigma ,\sigma '}\sum_{q}[A^{}_{\gamma\sigma;  \gamma'\sigma '}(\chi\tilde{\sigma}, E, E')\hat{a}^{\dagger}_{\gamma,\sigma}(E)\hat{a}^{}_{\gamma '\sigma '}(E')]\ .
\nonumber
\end{align}
The summation over the spin index $\tilde{\sigma}$ indicates that Eq. (\ref{IT}) pertains to the total electric current (as opposed to the spin-resolved one \cite{ENTIN}).
The indices $\gamma,\sigma$ ($\gamma ',\sigma '$) and the argument $E$ ($E'$) specify a scattering state of energy $E$ ($E'$) excited by an electron of spin polarization $\sigma$  ($\sigma '$) incoming from lead $\gamma$ ($\gamma '$) with $\gamma,\gamma '=L$ or $R$. \cite{ENTIN} The fermionic operators $\hat{a}^{\dagger}_{\gamma,\sigma }(E)$ and 
$\hat{a}^{}_{\gamma,\sigma }(E)$
create and annihilate an electron in the corresponding  scattering state. These operators are normalized such that
\begin{align}
\langle \hat{a}^{\dagger}_{\gamma\sigma}(E)\hat{a}^{}_{\gamma '\sigma '}(E')\rangle =\delta^{}_{\gamma\gamma '}\delta^{}_{\sigma\sigma '}\delta (E-E')f^{}_{\gamma}(E)\ .
\label{anor}
\end{align}
The matrix ${\bf A}$ is given in terms of the scattering matrix, $\mathbf{\cal S}$. \cite{BUTTIKER}  
In the tubular geometry considered here the transverse modes are not coupled, and therefore the scattering matrix splits into blocks of 4$\times$4
matrices for each value of the transverse momentum $q$ [see Eq. (\ref{Sfull});   we omit the argument $q$ from ${\bf A}$ for brevity]. 
For each value of $q$ the elements of the  (4$\times$4) matrix  ${\bf A}$  are given by
\begin{align}
A^{}_{\gamma \sigma ;\gamma '\sigma '}(\chi\tilde{\sigma},E,E')&=\delta^{}_{\gamma\gamma '}\delta^{}_{\sigma\sigma '}\delta_{\chi\gamma}\delta^{}_{\sigma\tilde{\sigma}}
\nonumber\\
&-{\cal S}^{\ast}_{\chi\tilde{\sigma};\gamma\sigma}(E){\cal S}^{}_{\chi\tilde{\sigma};\gamma '\sigma '}(E')\ , 
\label{A}
\end{align}
where $\mathbf{\cal S}$ is derived in Appendix \ref{App2} [see in particular Eq. (\ref{Sfull})].

The dc current through the scatterer is readily obtained by averaging Eq. (\ref{IT}) using Eq. 
(\ref{anor}). This leads to the celebrated Landauer formula \cite{IMRY} for the net current in terms of the transmission
\begin{align}
I&\equiv\langle\hat{ I}^{}_{L}\rangle=-\langle \hat{I}^{}_{R}\rangle \nonumber\\
&=\frac{e}{2\pi}\int_{-\infty}^{\infty}dE[f^{}_{L}(E)-f^{}_{R}(E)]{\cal T}(E)\ ,
\end{align}
where the transmission ${\cal T}$ is
\begin{align}
{\cal T}(E)=\sum_{q}{\cal T}^{}_{q}(E)=\sum_{q}{\rm Tr}[\mathbf{T}^{\dagger}_{LR,q}(E)\mathbf{T}^{}_{LR,q}(E)]\ .
\label{T}
\end{align}
Here the trace is carried out in spin space; the (2$\times$2) matrix $\mathbf{T}_{LR}$ (as well as the other matrices comprising the $q-$dependent scattering matrix) is given in Eqs. (\ref{RT}).  

As usual, we present the current correlations in the frequency domain, by defining \cite{IMRY}
\begin{align}
{\cal K}^{}_{\chi\chi '}(\omega )=\int_{-\infty}^{\infty} dt e^{i\omega t}\langle \delta\hat{I}_{\chi}(t)
\delta\hat{I}^{}_{\chi '}(0)\rangle\ ,
\end{align}
with $\delta \hat{I}_{\chi}(t)=\hat{I}_{\chi}(t)-\langle\hat{I}_{\chi}\rangle$. 
\begin{widetext}
Upon exploiting the relations (\ref{IT}) and (\ref{anor})
we find \cite{BUTTIKER}
\begin{align}
\label{KF}
&{\cal K}^{}_{\chi\chi '}(\omega )=
\frac{e^{2}}{2\pi}\int_{-\infty}^{\infty}dE\sum_{\gamma\gamma '}f^{}_{\gamma}(E)[1-f^{}_{\gamma '}(E+\omega )]
\sum_{q}
\sum_{\sigma\sigma '\tilde{\sigma}}A^{}_{\gamma\sigma;\gamma '\sigma '}(\chi\tilde{\sigma},E,E+\omega)A^{}_{\gamma '\sigma ';\gamma\sigma}(\chi '\tilde{\sigma},E+\omega, E)\ .
\end{align}
Inspecting Eq. (\ref{KF}), one can distinguish between two types of correlations,
the auto-correlations, for which $\chi=\chi '$, and the cross correlations, where $\chi\neq\chi '$.
For instance, when the setup is not biased, i.e., $\mu_{L}=\mu_{R}$, 
the auto-correlation  spectrum is given by
\begin{align}
\label{RR}
&{\cal K}_{RR}^{}(\omega )=\frac{e^{2}}{2\pi}\int_{-\infty}^{\infty}dEf(E)[1-f(E+\omega)]\sum_{q}{\rm Tr}[2
-\mathbf{R}^{\dagger}_{RR}(E)\mathbf{R}^{}_{RR}(E+\omega)-\mathbf{R}^{\dagger}_{RR}(E+\omega)\mathbf{R}^{}_{RR}(E)]\ ,
\end{align}
where the trace is carried out in spin space, and the (2$\times$2) matrix ${\bf R}$ is given in Eqs.   (\ref{RT}).
Likewise, the cross-correlation function for $\mu_{L}=\mu_{R}$ is
\begin{align}
\label{LR}
&{\cal K}_{LR}^{}(\omega )=\frac{e^{2}}{2\pi}\int_{-\infty}^{\infty}dEf(E)[1-f(E+\omega)]
\sum_{q}{\rm Tr}[\mathbf{T}^{\dagger}_{LR}(E)\mathbf{T}^{}_{LR}(E+\omega)+\mathbf{T}^{\dagger}_{LR}(E+\omega)\mathbf{T}^{}_{LR}(E)]\ .
\end{align}
When the junction is symmetric, i.e., the two tunnel junctions on  both its sides are of identical strength $\zeta_{L}=\zeta_{R}$ then ${\cal K}_{LL}(\omega )={\cal K}_{RR}(\omega )$
and ${\cal K}_{LR}(\omega )={\cal K}_{RL}(\omega )$. Otherwise, the expressions for ${\cal K}_{LL}$ and ${\cal K}_{RL}$ are obtained from Eqs. (\ref{RR}) and (\ref{LR}) upon interchanging 
$L$ with $R$.

The correlations of the physical quantities are combinations of the auto- and cross- correlations. For instance, since the operator of the net current  through the scatterer, $\hat{I}$,   
 reads 
\begin{align}
\hat{I}(t)=[\hat{I}^{}_{L}(t)-\hat{I}^{}_{R}(t)]/2\ ,
\end{align}
it is evident from Eq. (\ref{KF})
that the correlation of the net current is 
given by $[{\cal K}_{LL}+{\cal K}_{RR}-{\cal K}_{LR}-{\cal K}_{RL}]/4$. Likewise, the  charge correlation is \cite{ME}
 $[{\cal K}_{LL}+{\cal K}_{RR}+{\cal K}_{LR}+{\cal K}_{RL}]/4$.
In particular, in the zero-frequency limit  the net-current correlation is the shot noise, which is given by [see Eq. (\ref{KF})]
\begin{align}
\label{shot}
&{\cal K}^{}_{s}=
\frac{e^{2}}{2\pi}\int_{-\infty}^{\infty} dE\sum_{\gamma =L,R}\Big (f^{}_{\gamma}(E)[1-f^{}_{\gamma}(E)]{\rm Tr}\{{\cal T}(E)^{2}\}
+f^{}_{\gamma}(E)[1-f^{}_{\overline{\gamma}}(E)]{\rm Tr}\{{\cal T}(E)[1-{\cal T}(E)]\}\Big )\ ,
\end{align}
where $\overline{\gamma}$ is the lead opposite to the $\gamma $ lead. This result extends the celebrated expression 
first derived in Ref. \onlinecite{LESOVIK} to include the effects of spin-orbit interaction.

\end{widetext}

\subsection{Results}
\label{resfull}

\begin{figure}
  \includegraphics[width=6cm]{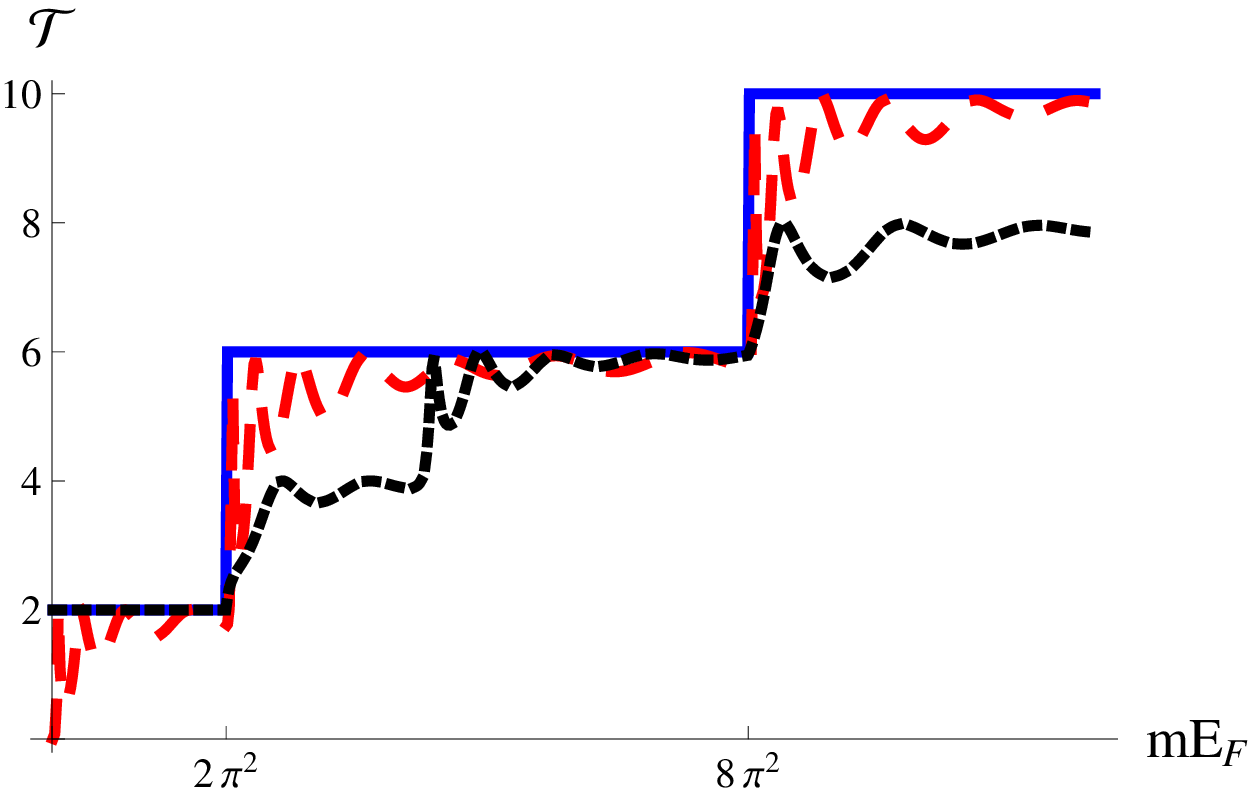}\\  \includegraphics[width=6cm]{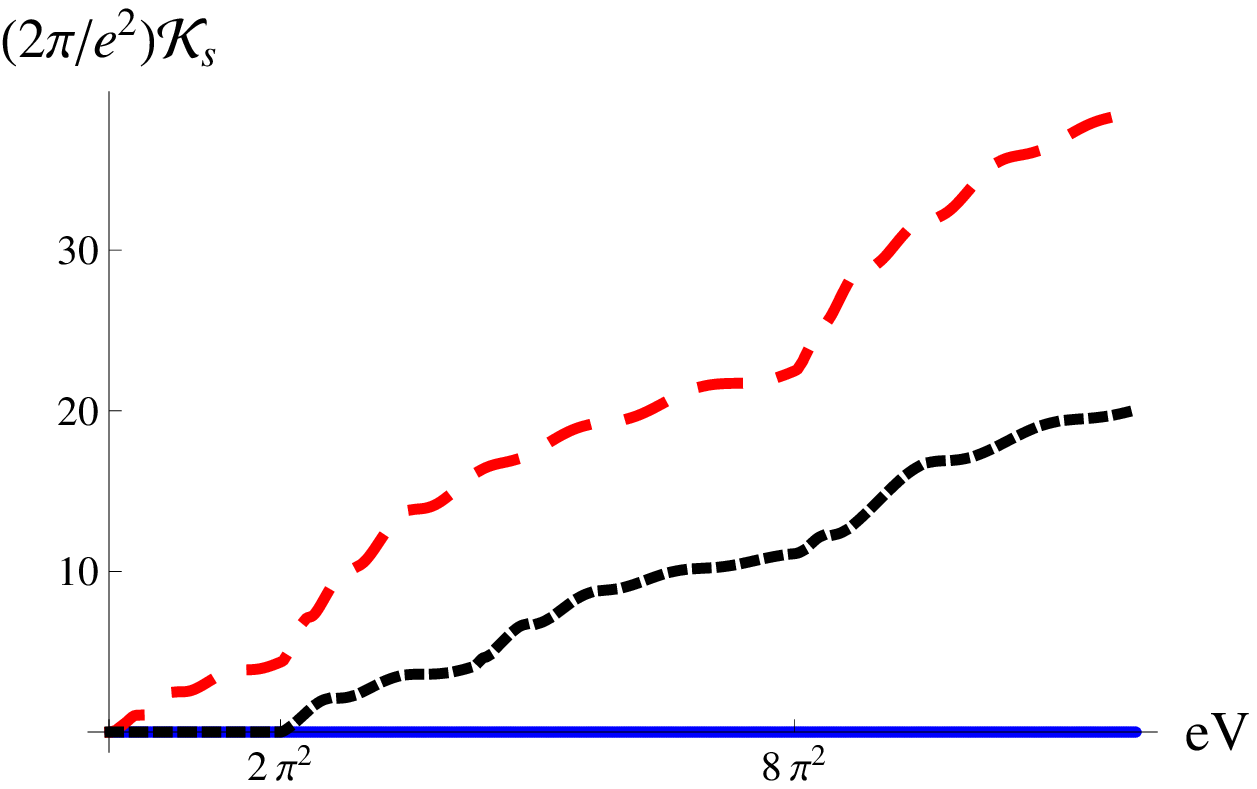} \\
  \includegraphics[width=6cm]{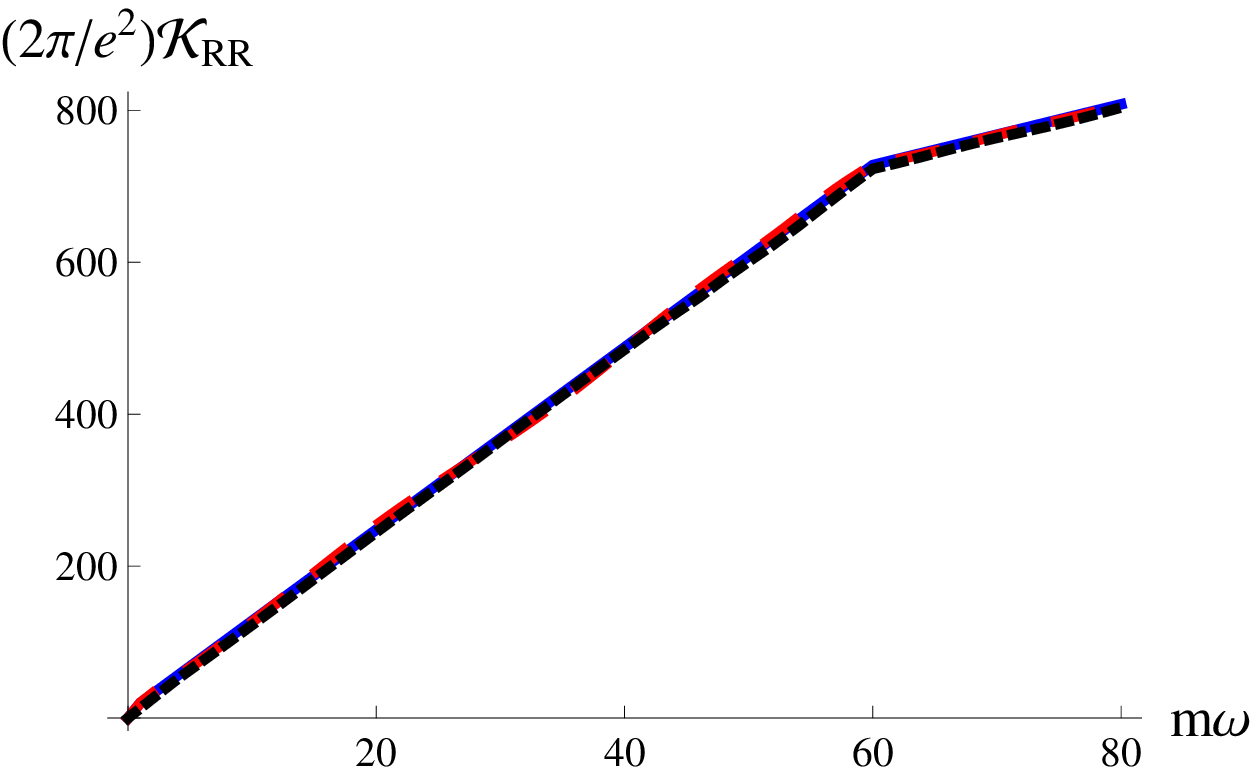}  
  \caption{(Color online) Top: the transmission as a function of the Fermi energy; mid panel: the shot noise Eq. (\ref{shot}) as a function of the bias voltage $e V=\mu_{L}-\mu_{R}$; bottom: the  auto-correlation Eq. (\ref{RR}) as a function of the external frequency (both the energy and the frequency are normalized by the mass $m$ and are measured in units of $L^{-2}$). The thin solid (blue) line in the top panel corresponds to the perfect conductor, in the absence of the spin-orbit coupling, the dotted (black) curve is for the case $\alpha =0.9 \pi$ and no potential barriers at the two ends, $\zeta_{L}=\zeta_{R}=0$, and the dashed (red)  curve  is for $\zeta_{L}=\zeta_{R}=1.2$, in units of $L^{-1}$. In the bottom panel $E_{F}=80$   in units of $m L^{-2}$; the solid (blue) and the dotted (black) curves there overlap.  
   }\label{FIG 1}
\end{figure}

In the absence of the spin-orbit interaction, the magnetic field, and the potential barriers at $x=\pm d$ (see Fig. \ref{figfull}), 
the transmission (\ref{T}) of the tubular 2DEG exhibits the well-known phenomenon of  perfect conductance quantization, whose hallmark is the staircase structure of the conductance (or the transmission) as a function of the Fermi energy (i.e., the gate voltage). 
Indeed, in this quintessential configuration the transmission amplitude matrices ${\bf T}_{LR}$ and ${\bf T}_{RL}$ [see Eqs. (\ref{RT})]
are both given by a unit matrix times $\exp[2id\sqrt{2mE_{}-q^{2}}]$, opening a new transverse channel whenever $E_{}$ is large enough for an additional  $q$ to yield a real $k$ [see Eq. (\ref{k})]. Note though, that as opposed to a flat two-dimensional wire, here the quantization steps appear for $n=2,$ 6, 10, {\it etc.} \cite{ENTIN} [see Eq. (\ref{q})] reflecting the helical degeneracy of the $q$ values for the cylinder. 
Since the conductor is perfectly transmitting, the shot noise Eq. (\ref{shot}) vanishes;  the auto- and cross- correlations do not. For instance,  at zero temperature and for an un-biased system, $\mu_{L}=\mu_{R}=E_{F}$,   Eqs. (\ref{RR}) and (\ref{LR}) yield
\begin{align}
&\frac{\pi}{2e^{2}}{\cal K}^{}_{RR}(\omega )=\omega\ ,\ \ {\rm per}\ \ {\rm channel}\ ,\nonumber\\
&\frac{\pi}{2e^{2}}{\cal K}^{}_{LR}(\omega )=\int_{E^{}_{F}-\omega}^{E^{}_{F}}dE\sum_{q}\nonumber\\
&\times\cos[2d(\sqrt{2mE-q^{2}}-\sqrt{2m(E+\omega)-q^{2}})]\ .
\end{align}
The fact that the current correlations do not vanish for an un-biased conductor at zero temperature reflects the relation between the noise spectrum and the absorption/emission capacity of the scattering system. \cite{AGUADO,GAVISH} These results are exemplified in Fig. \ref{FIG 1} by the  solid (blue) curves.

The spin-orbit interaction lifts partially   the helical degeneracy.
As mentioned, one of the longitudinal wave vectors, $k_{\rm d}$,  can become imaginary. This happens when $2mE<(q+\alpha)^{2}$; in that case the corresponding wave is evanescent and does not contribute to the transmission [recall that energy is measured from $\alpha^{2}/(2m)$].
As a result, the conductance steps are split \cite{ENTIN} (save the 
first one, which, as explained above, is insensitive to the spin-orbit interaction).
Moreover, since the transmission is not perfect anymore, the shot noise (as a function of the bias) is finite. 
As can be seen in
the mid panel of Fig. \ref{FIG 1}, the shot noise 
begins  at   small bias voltages $V$ with a nearly horizontal slope, corresponding to a fully-transmitting channel. It then develops 
a steeper slope, reflecting the partially-transmitting channel ({\it cf.} the top panel). The fact that each stair (save the very first one) is split into two means that in the low-energy part of the stair only one of the spinors is propagating (the other belongs to the evanescent wave); in other words, the electric current is spin polarized. \cite{ENTIN}

\begin{figure}
     \includegraphics[width=6cm]{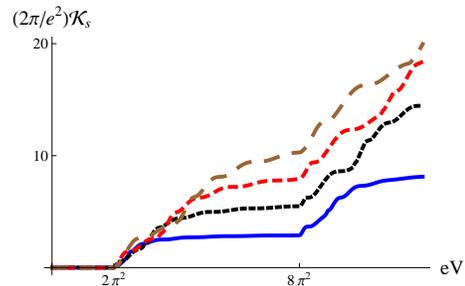} 
  \caption{(Color online) The shot noise as a function of   bias voltage, for a completely open cylinder, $\zeta_{L}=\zeta_{R}=0$.  The solid  (blue) curve is for $\alpha=0.2\pi$, the dotted (black) one is for  
  $0.4\pi$, the small-dashed  (red) line is for    $0.6\pi$, and the large-dashed (brown) line is for  
   $0.8\pi$; all  in units of $L^{-1}$.}
  \label{FIG SHOTA}
\end{figure}

The effect of the spin-orbit coupling on the shot noise is better appreciated from Fig. \ref{FIG SHOTA}, where it is plotted (as a function of the bias voltage) for various values of the coupling strength $\alpha$ (in units of inverse length). Grossly speaking, the staircase structure is gradually lost as the spin-orbit coupling increases. Perhaps the main feature of the shot noise brought about by the spin-orbit interaction is the  division between regions in which it is roughly horizontal and where it is approximately linearly increasing.  The first pertains to the case in which both spinors are transmitted, while the second describes the situation where one of the spinors is blocked. In this way, the shot noise may serve as an indicator for a spin-polarized current.

\begin{figure}
   \ \includegraphics[width=7cm]{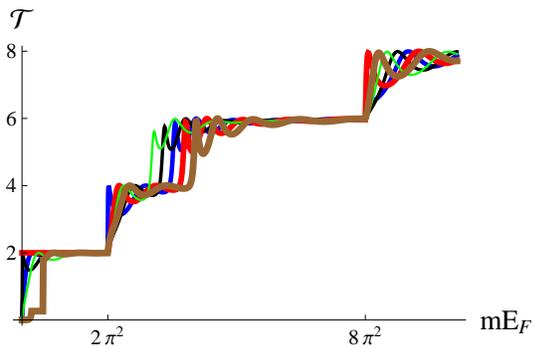}   
  \caption{(Color online) The transmission as a function of   the Fermi energy. The various curves are for different values of the gate voltage $U$    (see text)  . In increasing order of the thickness, the thinnest (blue) curve is  for $U=0$, then ,  $mU=-(0.7\pi)^2/2$   (black), $mU=(0.7\pi)^2/2$  (red), $mU=-(0.7\pi)^2$ (green), and the thickest $mU=(0.7\pi)^2$ (brown), in units of $L^{-2}$. Here  $\alpha=0.7\pi$, and $\zeta_{L}=\zeta_{R}=0$.  
  }\label{FIG 2}
\end{figure}

It is also of interest to explore the effect of a gate voltage applied uniformly on the scatterer. To this end we add to the Hamiltonian (\ref{HAMFUL})
the term $U\Theta (d-|x|)$. The gate potential $U$
which acts as a potential well/wall modifies the width of the conductance stairs. As mentioned above, the spin-orbit interaction splits each conductance/transmission stair into two; the width (in energy) of the split step 
is $q^{2}-2\alpha|q|\leq 2m (E_{F}-U)\leq q^{2}+2\alpha |q|$, for $2mU\geq 2\alpha|q|$. This behavior is depicted in Fig. \ref{FIG 2}.
The width may then be controlled by the gate voltage; in view of the comments  above (in conjunction with Fig. \ref{FIG 1})
we conclude that by varying the gate voltage one may manipulate the spin polarization of the electric current.

\begin{figure}
   \includegraphics[width=7cm]{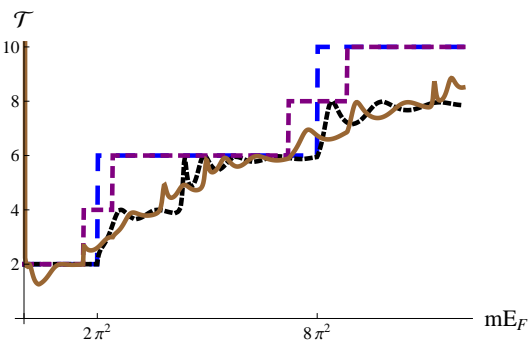}    \\
  \caption{(Color online) The transmission as a function of  the Fermi energy. The largest-dashed and the dotted curves   (blue and black) are the same as in Fig. \ref{FIG 1}, and are reproduced here as a reference. The solid and small-dashed  lines (brown  and purple) exhibit the effect of a magnetic flux (we use $\varphi=0.2\pi$). The staircase one (purple) is for $\alpha=0$, the wavy one (brown) is for  $\alpha=0.9\pi$ in units of $L^{-1}$.}
  \label{FIG 5}
\end{figure}

Another  tool to lift the helical degeneracy is to apply a magnetic flux along the tube axis.
In the absence of the flux,  the scattering states for $n$ and for $-n$ are degenerate [$n$ is the quantum number   of the transverse modes, see Eq. (\ref{q})]. The magnetic flux removes 
this degeneracy, as illustrated in Fig. \ref{FIG 5}, turning the 4-fold degeneracy into a two-fold one.
The reason being the modification of the transverse wave vector $q$ by the relative flux $\varphi$,  $q\rightarrow q-\varphi$, as discussed after Eq. (\ref{HAMFUL}).
Thus for example, the threshold for the opening of a new step in the transmission is  $2mE_{F}>{\rm min}(q-\varphi)^{2}$, where ``min" stands for the minimal value of $(q-\varphi)^{2}$ for all $q-$values. 
Under the action of both a magnetic field and the spin-orbit coupling, the  entire 4-fold degeneracy is removed, as  shown by the thick, very wavy  (brown) curve in Fig. \ref{FIG 5}.
It follows that  manipulating the gate voltage and  the magnetic field in a Rashba scatterer enables a good control on  both the helicity and the spin degrees of freedom of the transmitted electrons.

\begin{figure}
  \includegraphics[width=5cm]{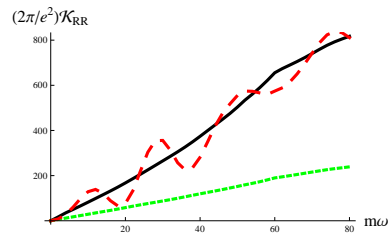} 
  \caption{(Color online) The auto-correlation noise, ${\cal K}_{RR}(\omega)$, as a function of  frequency (normalized by the mass $m$ and measured in units of $L^{-2}$), for $\alpha =3.1 \pi$ (in units of $L^{-1}$). The solid (black) reference  curve corresponds to $\zeta_{L}=\zeta_{R}=0$, the dotted (green) line is for $\zeta_L=1.2$ and $\zeta_R=19.2$, while the dashed (red) line is  for $\zeta_L=19.2$ and $\zeta_R=1.2$, all  in units of $L^{-1}$.}\label{FIG 6}
\end{figure}

We now turn to discuss the current-correlation functions [see e.g., Eqs. (\ref{RR}) and (\ref{LR})]. These are known to be rather sensitive to asymmetries of the setup, in our case to a possible difference  between $\zeta_{L}$ and $\zeta_{R}$. \cite{ME,GABDANK}
Figure \ref{FIG 6} displays the auto-correlation ${\cal K}_{RR}$, in the absence of the spin-orbit coupling and a bias voltage.
There is a  distinct disparity between ${\cal K}_{RR}$ pertaining to the case where  the left tunnel junction is almost pinched off (the wavy line) and  when it is almost open (the lower thick line; the thin line is for $\zeta_{L}=\zeta_{R}=0$,  and serves as a reference) where the noise is considerably lower.
The auto-correlation decreases as $\zeta_{R}$ increases, and vanishes when this tunnel junction is pinched off.

\begin{figure}
      \includegraphics[width=2.8in]{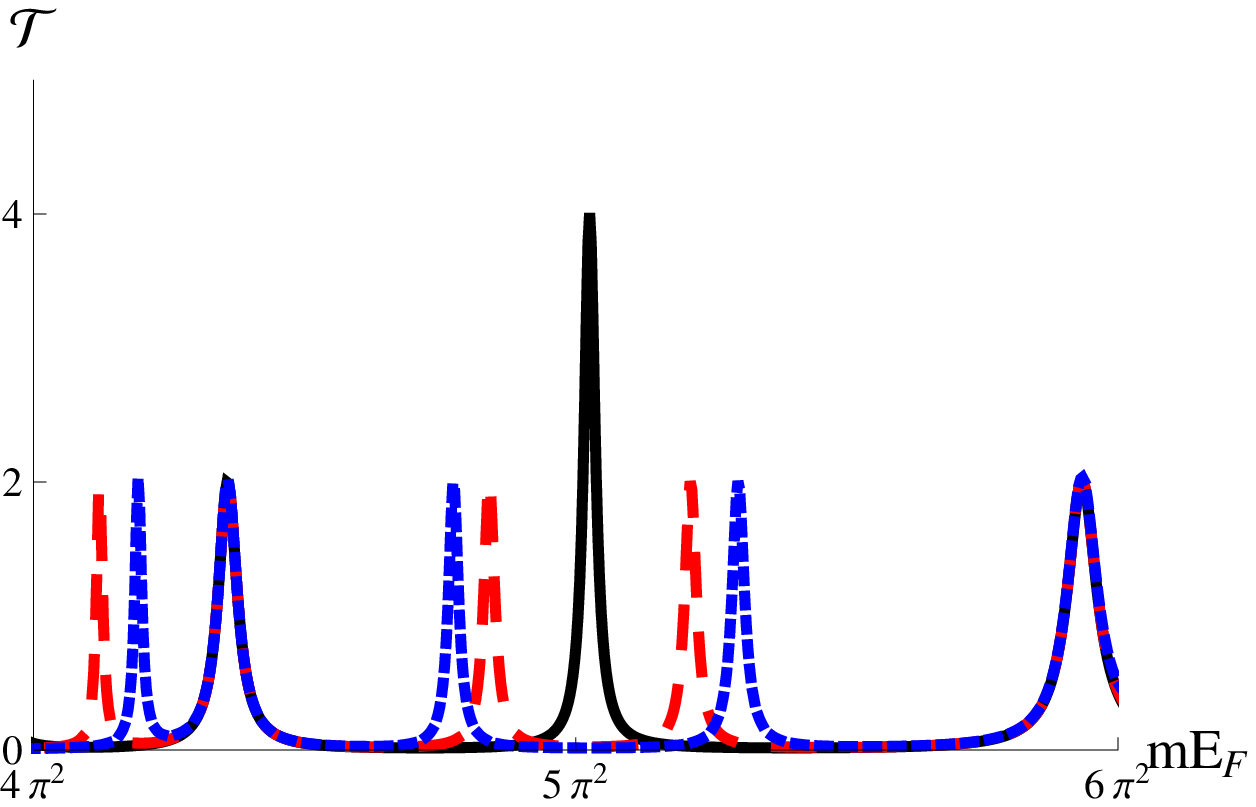}\\
        \includegraphics[width=2.8in]{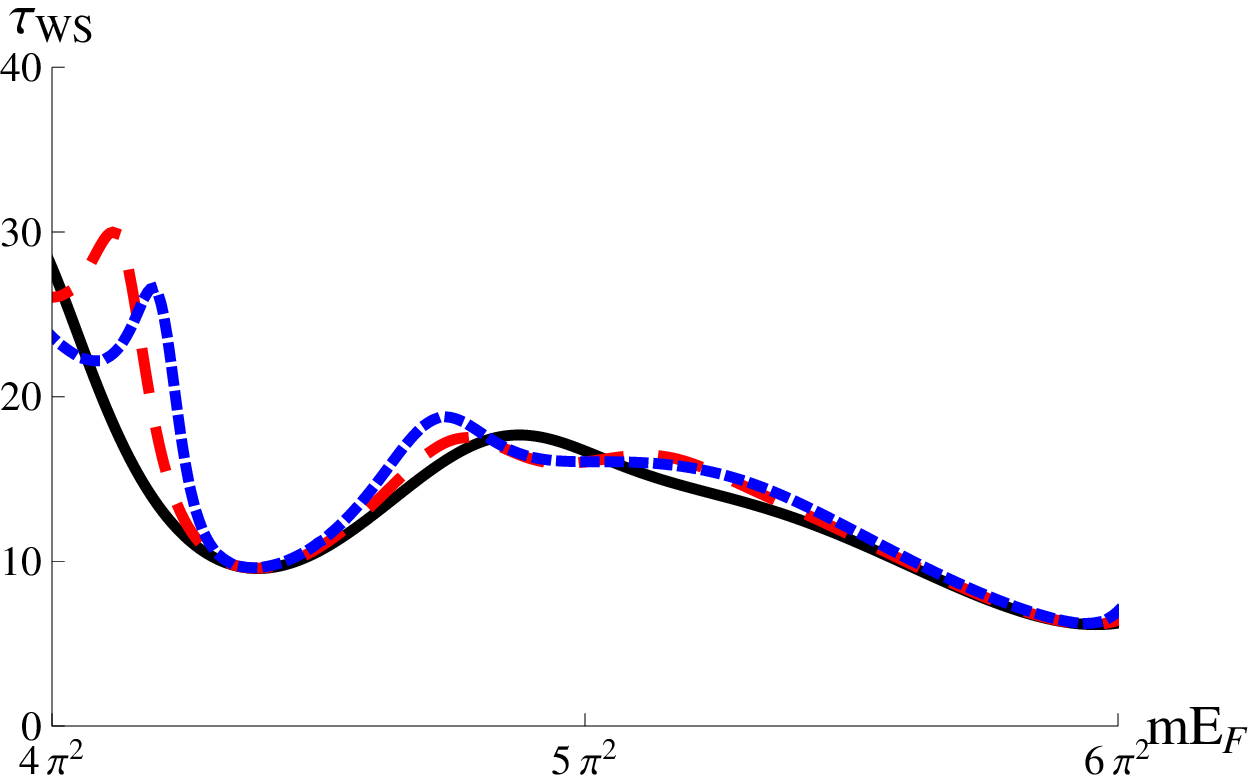} \\
  \caption{(Color online) The upper panel displays the transmission as a function of the Fermi energy (in unites of $L^{-2}$) and the lower panel shows the  Wigner-Smith time (in units of $\hbar$), again as a function of the Fermi energy.
 Here $d=1$ (in units of $L$), $\phi=0$,  and $\zeta=30$ (in units of $L^{-1}$). The solid (black) curve is for $\alpha=0$, the dotted one  (blue) one   is for $\alpha=0.2\pi$, and  the dashed line (red) curve is for $\alpha=0.1\pi$.   }
 \label{FIG10}
\end{figure}

\subsection{The Wigner-Smith  time-delay  matrix }

\label{QF}

We conclude this section with a discussion of the Wigner-Smith  time-delay matrix of  our core-shell structure. 
Smith \cite{Smith} introduced  the time-delay hermitian matrix
\begin{align}
\tau^{}_{\gamma\gamma '}(E_{F}^{})=
\sum_{q}{\rm Tr}
\Big (\frac{1}{2\pi i}\mathbf{S}^{\dagger}_{\gamma \gamma '}(E)\frac{d\mathbf{S}^{}_{\gamma\gamma '}(E)}{dE}\Big )\Big |^{}_{E_{F}}\  , 
\end{align}
whose diagonal matrix elements measure the average lifetime of a scattering  event (the collision lifetime in the terminology of Ref. \onlinecite{Smith}). Here
$\gamma, \gamma '=L,R$,  and $\mathbf{S}$ is  the scattering matrix pertaining 
to a certain value of $q$ (see Appendix \ref{App2}).
Explicitly, $\tau_{\gamma\gamma '}(E_{F})$ is the time delay experienced by an electron at the Fermi energy
incident from the  $\gamma$ lead  into the  $\gamma '$ one   (including in our case all  transverse channels and the spin polarizations) because of the scatterer. This quantity is intimately related to the quantum capacitance studied in the previous section, 
and is also related to the thermopower the scatterer is capable of producing. \cite{van Langen}
Indeed, 
by 
expanding the dynamic conductance \cite{BUTTIKER3} ${\cal G}_{\gamma \gamma '}(\omega )$
\begin{align}
&{\cal G}^{}_{\gamma\gamma '}(\omega)=\frac{e^{2}}{2\pi}
\int\frac{dE}{\omega}[f(E)-f(E+\omega )]\nonumber\\
&\times\sum_{q}{\rm Tr}[\delta^{}_{\gamma\gamma '}-\mathbf{S}^{\dagger}_{\gamma\gamma '}(E)\mathbf{S}^{}_{\gamma \gamma'}(E+\omega )]\ ,
\end{align}
to first order in the external frequency $\omega$, one finds that at zero temperature
\begin{align}
{\cal G}^{}_{\gamma \gamma '}(\omega )\simeq{\cal G}^{}_{\gamma \gamma '} (0)
-ie^{2}\omega\tau^{}_{\gamma \gamma '}\ .
\label{secex}
\end{align}
For
$\gamma \neq \gamma '$, e.g., $\gamma =L$ and $\gamma '=R$, the
first term on the right-hand side of Eq. (\ref{secex}) is the transmission given in Eq. (\ref{T}) times the quantum unit of the conductance [$e^{2}/(2\pi)$ for a single spin, with $\hbar=1$], i.e., the Landauer conductance. The simple separation of the ac conductance into  real and imaginary parts which appears in Eq. (\ref{secex}) led to the identification of
the quality factor of the mesoscopic conductor as roughly the  absolute value of the ratio  ${\rm Im}{\cal G}/{\rm Re}{\cal G}$. 
\cite{RAHACHOU2} (Reference \onlinecite{RAHACHOU2} replaces the denominator  by the number of channels up to the Fermi energy. This replacement is apparently valid when the transmission is close to resonance.)
The quality factor
measures the capability of a circuit to store energy; its enhanced value in carbon nanotubes is a subject of current interest. \cite{HUTTEL} 
Our analysis below is carried out for an  ``average delay time", $\tau_{WS}$, defined by
\begin{align}
\tau^{}_{WS}=\frac{2\pi}{\cal T}
\sum_{\gamma, \gamma '}\tau^{}_{\gamma\gamma '}\ .
\label{tws}
\end{align}
Note that $\tau_{WS}$ is measured in units of $\hbar$.

\begin{figure}
        \includegraphics[width=2.8in]{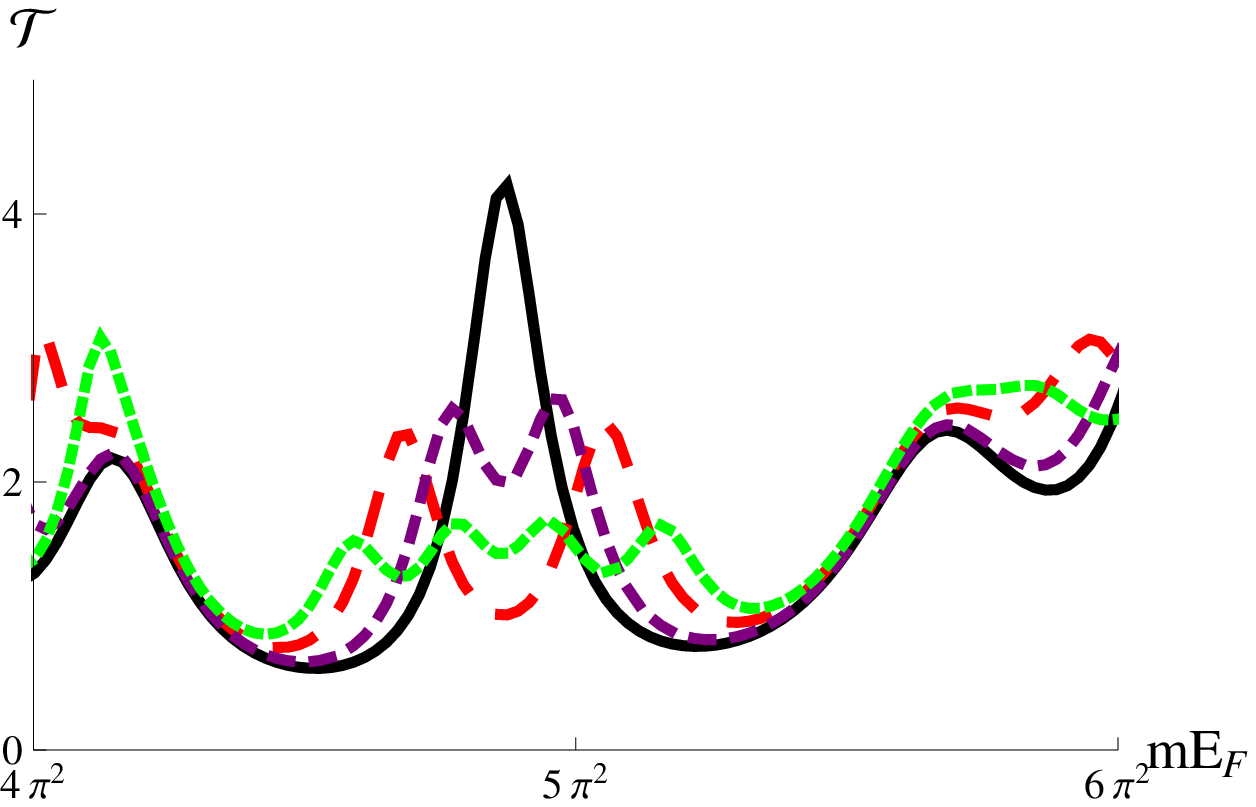} \\
            \includegraphics[width=3in]{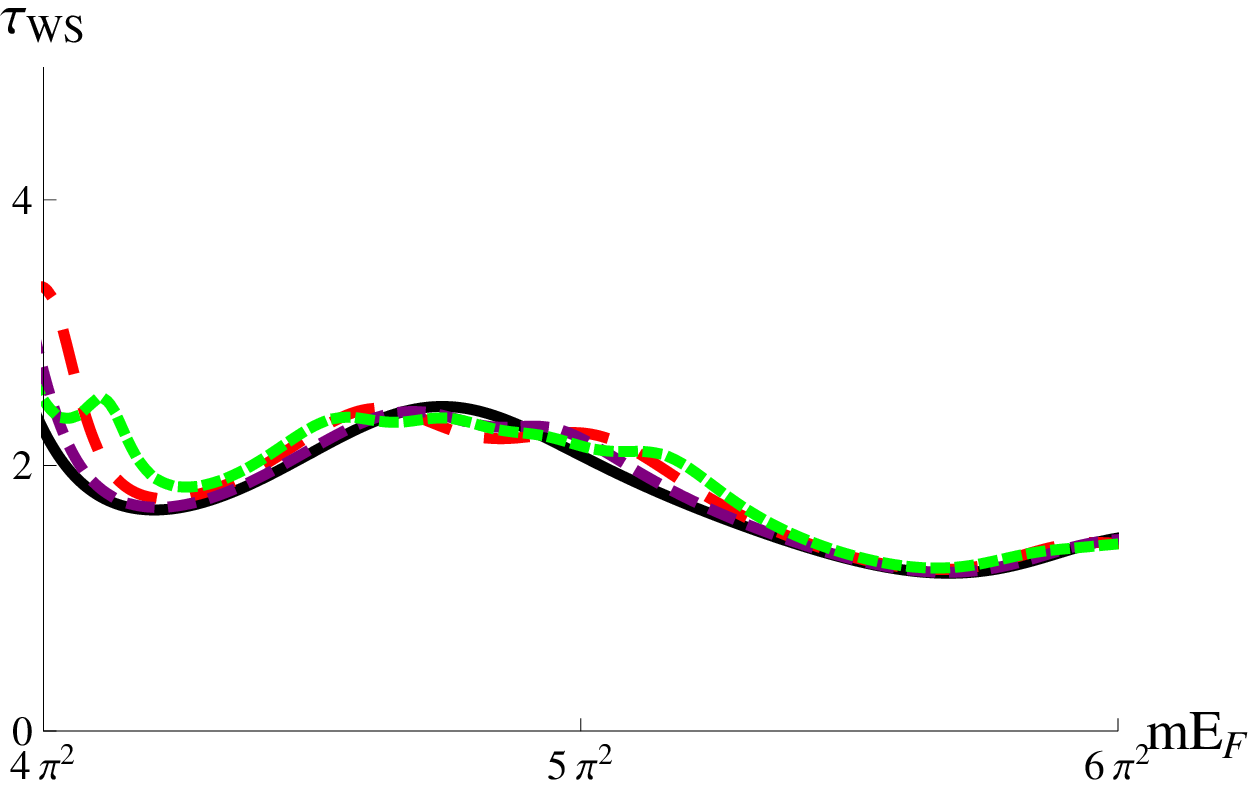} \\  \caption{(Color online) The upper panel is the transmission and the lower panel  is $\tau_{WS}$, both  as a function of the Fermi energy.  Here $d=1$ (in units of $L$) and $\zeta=9.7$ (in units of $L^{-1}$. The solid (black) curve is for $\alpha=\phi=0$, the small-dashed  (purple)  curve is for $\alpha=0.1\pi$ and $\phi=0$, the large-dashed curve   (red)  is for $\alpha=0$ and $\phi=0.05\pi$, and the dotted (green) one  is for $\alpha=0.1\pi$ and $\phi=0.05\pi$.}
            \label{FIG11}
\end{figure}

As explained by Smith,  \cite{Smith} close to resonance  the delay time is related to the (inverse of the) resonance width.  In fact, when the scattering matrix  can be described by  a simple Breit-Wigner resonance of width $\Gamma$, $\tau_{WS}$ as defined in Eq. (\ref{tws}) equals $\Gamma^{-1}$.  In an attempt to investigate this feature in a core-shell system, we confine ourselves in this subsection to high tunnel barriers such that the transmission consists of narrow peaks. (For simplicity a symmetric setup where $\zeta_{L}=\zeta_{R}\equiv \zeta$ is considered.)
Figures 
\ref{FIG10}  and \ref{FIG11} display  (in the upper panels) the transmission as a function of the Fermi energy around one of the (relatively) sharp peaks  (the higher is $\zeta$, the sharper are the transmission peaks), and in the lower panels the corresponding  Wigner-Smith time, Eq. (\ref{tws}).  As can be expected, the average delay time does vary with the Fermi energy, though the transmission is close to resonance. An example is shown in Fig. \ref{FIG10}. The full (black) curves in the two panels there are  the transmission and $\tau_{WS}$ in the absence of the spin-orbit coupling. It is rather straightforward to find that  for $\alpha=0$
\begin{align}
{\cal T}=\sum_{q}\Big (1+4\frac{\zeta^{2}}{k^{2}}[k\cos(2kd)+\zeta \sin(2kd)]^{2}\Big )^{-1}\ ,
\end{align}
where $q$ and $k$ are given by Eqs. (\ref{q}) and (\ref{k}), respectively, and
\begin{widetext}
\begin{align}
\tau^{}_{WS}=\frac{1}{\cal T}\sum_{q}\frac{8d}{v(E_{F})}\frac{1+\frac{\zeta}{k^{2}d}+\frac{2\zeta^{2}}{k^{2}}+\frac{2\zeta^{2}}{k^{4}d}\sin (2kd)[\zeta\sin(2kd)+k\cos (2kd) ]}{1+4\frac{\zeta^{2}}{k^{2}}[k\cos(2kd)+\zeta \sin(2kd)]^{2}}\ , 
\label{TWS}
\end{align}
\end{widetext}
where $v(E_{F})\equiv k/m$ is the velocity of the electron at  the Fermi energy. For instance, when $\zeta=0$
Eq. (\ref{TWS}) gives  for $\tau_{WS}$ the value $4\times  2d/v(E_{F})$, which is the  time required  for an electron to traverse ballistically  a  tube of length 2$d$, times the spin and helical degeneracies.

Figure \ref{FIG10} displays the  Wigner-Smith time for a rather sharp transmission peak. The curves are for different values of the spin-orbit coupling; it is seen that while this coupling has a substantial effect on the transmission by removing the spin degeneracy (splitting the peak into two), it hardly changes $\tau_{WS}$. The same feature can be observed in Fig. \ref{FIG11}; there we have added the effect of the magnetic flux, which lifts the helical degeneracy. Nonetheless, the Wigner-Smith time is almost unchanged.  
Comparing the two figures,   \ref{FIG10} and \ref{FIG11}, 
it is observed that (not surprisingly) $\tau_{WS}$ increases significantly  as the  transmission peak
becomes narrower.


\section{Conclusions}
\label{conclu}

We have investigated several frequency-dependent properties of a tubular two-dimensional electron gas, subject to the Rashba spin-orbit interaction. In this  quintessential  geometry the spin-orbit coupling does not mix the transverse channels, and therefore the effect of the interaction can be related 
 to interference.

 We have found that when the tubular two-dimensional electron gas is coupled to a single reservoir the quantum capacitance ${\cal C}$ and the charge-relaxation resistance ${\cal R}$
 are sensitive probes of the charging state of the scatterer. In particular peaks in ${\cal C}$ correspond to resonance states (standing waves of the isolated segment) while peaks in ${\cal R}$ correspond to minima of ${\cal C}$ [as expected from Eqs. (\ref{c}) and (\ref{r})] in between the peaks of ${\cal C}$ or at zeroes of ${\cal C}$ (see Figs. \ref{FIG AZ}); an exception is the $q=0$ case where 
${\cal R}$ attains the universal value  $\pi/(2e^{2})$.

For the tubular system with two reservoirs we find that  the interference related to the spin-orbit coupling is  in particular manifested in the upper panel of Fig. \ref{FIG11}, where we see the similar manner by which both an orbital magnetic field (directed along the axis of the tube) and the spin-orbit coupling (that can also be assigned an effective magnetic field; the interference effect of the spin-orbit interaction is mainly due to the effective field associated with the transverse direction,  i.e., along the $\hat{\bf x}-$direction) affect the transmission. 
 The other remarkable effect of the spin-orbit coupling is its capability to block one of the propagating spinors in each transverse channel. \cite{ENTIN,EAV} In the case of the transmission, this is translated into  splitting of the  stairs (as a function of the Fermi energy, and for not too strong $\alpha$). In the case of the shot noise, this blocking modifies the shot noise as a function of the bias voltage. Since when one of the propagating spinors is blocked the current  is spin-polarized, measurements of the conductance and the shot noise can indicate the range of Fermi energies where this polarization takes place.

\begin{acknowledgments}
This work was supported by the Israeli Science Foundation (ISF), by
the Infrastructure program of Israeli Ministry of Science and Technology under the contract 3-11173, 
and the US-Israel Binational Science Foundation (BSF).
 \end{acknowledgments}

\appendix

\section{The reflection matrix  (Fig. \ref{figrc})}
\label{App1}

The scattering matrix for the geometry of Fig. \ref{figrc} is reduced to  a reflection matrix; for instance,  in the absence of the spin-orbit interaction and the delta-function potential at the interface $x=-d$ it is proportional to the unit matrix (suitably choosing  the origin).
In order to derive it, we write for the wave function the Ansatz
\begin{align}
\varphi^{}_{b}(x,y)&=\sum_{q}e^{iqy}(e^{ikx}|c^{}_{\rm in}\rangle +e^{-ikx}|c^{}_{\rm out}\rangle )\ ,\label{phib}
\end{align}
in the ballistic region $x\leq -d$, and
\begin{align}
\varphi^{}_{t}(x,y)=\sum_{q}
e^{iqy} ({\bf v}^{}_{\rm u}e^{ik^{}_{\rm u}x\sigma^{}_{z}}|c^{}_{\rm u}\rangle +
{\bf v}^{}_{\rm d}e^{ik^{}_{\rm d}x\sigma^{}_{z}}|c^{}_{\rm d}\rangle)\ ,
\label{phit}
\end{align}
in the mesoscopic conductor, in the region $-d\leq x\leq 0$. Here $\sigma_{z}$ is the third Pauli matrix.  Equation
(\ref{phit}) is valid for positive values of $q$,  Eq. (\ref{q}); a similar calculation holds for the negative values.
The matrices $\mathbf{v_u}$ and $\mathbf{v_d}$ are
\begin{align}
\mathbf{v_u}&=\left [
                    \begin{array}{cc}
                      v^{}_{\rm u} & -v_{\rm u}^{-1} \\
                      -v_{\rm u}^{-1} & v^{}_{\rm u} \\
                    \end{array}
                  \right ]\ ,\ \ \ \ \ \ 
\mathbf{v_d}=\left [
                    \begin{array}{cc}
                      v^{}_{\rm d} & v_{\rm d}^{-1} \\
                      v_{\rm d}^{-1} & v^{}_{\rm d} \\
                    \end{array}
                  \right ]\ ,
\label{V}
\end{align}
with 
\begin{align}
v^{}_{\rm u,d}=\left(\frac{q+ik^{}_{\rm u,d}}{q-ik^{}_{\rm u,d}} \right)^{\frac{1}{4}}\ .
\label{v}
\end{align}
The various coefficients in Eqs. (\ref{phib}) and (\ref{phit}) are obtained by imposing boundary conditions. Explicitly
\begin{align}
{\bf v}^{}_{\rm u}|c^{}_{\rm u}\rangle +{\bf v}^{}_{\rm d}|c^{}_{\rm d}\rangle =0\ ,
\label{bc0}
\end{align}
to ensure the vanishing of $\varphi_{t}$ at $x=0$, 
and
\begin{align}
&\varphi^{}_{b}(-d,y)=\varphi^{}_{t}(-d,y)\ ,\nonumber\\
&\Big (\frac{\partial \varphi^{}_{b}(x,y)}{\partial x}-\frac{\partial \varphi^{}_{t}(x,y)}{\partial x}\Big )\Big |_{x=-d}^{}\nonumber\\
&=2\zeta \varphi^{}_{b}(-d,y)+i\alpha\sigma^{}_{y}\varphi^{}_{b}(-d,y)\ ,
\label{bcd}
\end{align}
for continuity at $x=-d$. Using Eqs. (\ref{bc0}) and (\ref{bcd}) to eliminate the vectors
$|c_{\rm u,d}\rangle$
results in the relation
\begin{align}
e^{-ikd}|c^{}_{\rm in}\rangle ={\bf F}(e^{-ikd}|c^{}_{\rm in}\rangle +e^{ikd}|c^{}_{\rm out}\rangle )\ ,
\label{SM1}
\end{align}
\begin{align}
{\bf F}=F^{}_{1}+\sigma^{}_{x}F^{}_{x}+\sigma^{}_{z}F^{}_{z}\ ,
\label{F}
\end{align}
with
\begin{widetext}
\begin{align}
F^{}_{1}&=
\frac{1}{2}+\frac{i \zeta }{k}
+\frac{i}{\Omega}   \Big (k_{\rm d}^{} \cos \left(d k^{}_{\rm d}\right) \sin \left(d k^{}_{\rm u}\right) [2  \alpha  \sqrt{2m E }+4mE] +k^{}_{\rm u} \sin
   \left(d k^{}_{\rm d}\right) \cos \left(d k^{}_{\rm u}\right) [4 mE-2 \alpha  \sqrt{2m E }]\Big)\ ,\nonumber\\
F^{}_x&=\frac{2iq \sqrt{2m E } }{\Omega}\Big (k^{}_{\rm d} \cos \left(d k^{}_{\rm d}\right) \sin \left(d k^{}_{\rm u}\right)-k^{}_{\rm u} \sin \left(d
k^{}_{\rm d}\right) \cos \left(d k^{}_{\rm u}\right)\Big ) \ ,\nonumber\\
F^{}_z &=\frac{iq}{\Omega}\Big (2 [k^2+\alpha ^2]\sin \left(d k^{}_{\rm d}\right) \sin \left(d k^{}_{\rm u}\right)+2 k^{}_{\rm d} k^{}_{\rm u} [\cos \left(d k^{}_{\rm d}\right) \cos \left(d
k^{}_{\rm u}\right)-1]\Big )\ .
\label{FC}
\end{align}
Here we have defined
\begin{align}
\Omega=4 k \Big ([2 m E +q^2-\alpha ^2]\sin \left(d k^{}_{\rm u}\right)\sin \left(d k^{}_{\rm d}\right) +k^{}_{\rm d} k^{}_{\rm u} [1-\cos
 \left(d k^{}_{\rm d}\right) \cos \left(d k^{}_{\rm u}\right)]\Big )\ .
 \label{OM}
\end{align}
\end{widetext}
Finally we note that in the first transverse channel, i.e., for $q=0$, $F_{x}$ and $F_{z}$ vanish, while
\begin{align}
F_1= \frac{1}{2}+\frac{i \zeta }{k} +\frac{i \left(k^{}_{\rm u}+k^{}_{\rm d}\right) \cot [\frac{1}{2}\left(k^{}_{\rm u}+k^{}_{
\rm d}\right)d]}{4 k}\ .
\end{align}
Note that in order to obtain Eq. (\ref{SM}) from Eq. (\ref{SM1})
we have shifted the origin by $-d$, i.e., $\exp[ikd ]|c_{\rm out}\rangle\rightarrow
|c_{\rm out}\rangle$ and
$\exp[-ikd ]|c_{\rm in}\rangle\rightarrow
|c_{\rm in}\rangle$.

As mentioned in the text,  $k_{\rm d}$ can be either real or imaginary. However, in both cases  $F_x$ and  $F_z$ are purely imaginary, and $F^{}_{1}+F_{1}^{\ast}=1$.  These  properties ensure that the reflection matrix ${\bf R}(q)$ [see Eq.  (\ref{SM})] is unitary. [The condition ${\bf F}_{}+{\bf F}^{\dagger}_{}=1$ (for each value of $q$)  is dictated by the optical theorem.]  A straightforward algebra gives that for each value of $q$ the reflection matrix is given by
\begin{align}
{\bf R}&=\sqrt{\frac{[F^{\ast}_{1}]^{2}-F^{2}_{x}-F^{2}_{z}}{F^{2}_{1}-F^{2}_{x}-F^{2}_{z}}}
\nonumber\\
&\times\frac{|F^{}_{1}|^{2}+F^{2}_{x}+F^{2}_{z}-\sig\cdot(F^{}_{x},0,F^{}_{z})}{\sqrt{(|F^{}_{1}|^{2}+F^{2}_{x}+F^{2}_{z})^{2}-F^{2}_{x}-F^{2}_{z}}}\ ,
\label{RAP}
\end{align}
which leads to the form (\ref{SSM}) used in Sec. \ref{rc}.

We have chosen ${\bf R}$ to define the reflection at $x=-d$. This choice, which is not so benign, reflects our expectation that the  displacement of the electrons [see Eq. (\ref{sumr})] is confined to the range $-d<x<0$ so that the quantum capacitance, Eq. (\ref{DOS}), will be positive.   We have seen, however, that very close to the singular steps this may not be the case.  Negative capacitance has been given various interpretations \cite{JONSCHER}, while here it reflects the spatial range of charge response. Ideally one could measure the whole wire and then the capacitance  would be always positive and Eq. (\ref{DOS}) would become exact.

We end this appendix by listing the explicit analytical expressions for the scattering phases 
for the case where the spin-orbit coupling vanishes.
Then the eigenvalues of the matrix $\mathbf{F}$, Eq. (\ref{lam}), and the corresponding phases Eq. (\ref{ep}),   are
\begin{align}
  \lambda^{}_{1}= \lambda^{}_{2}=\frac{1}{2}+i\frac{\zeta}{k}+\frac{i}{2}\cot k\ ,
\end{align}
and
\begin{align}
\label{eigen zeta}
  e^{i\phi^{}_{1}}= e^{i\phi^{}_{2}} =e^{2ikd}\frac{\zeta  \left(1-e^{-2 i d k}\right)+i k }{\zeta  \left(1-e^{2 i d k}\right)-i k }\ , 
\end{align}
with 
\begin{align}
&\frac{\partial\phi^{}_{1}}{\partial E}=
\frac{m}{k}\Big (2d\nonumber\\
&+\frac{-4\zeta \sin(kd)[(2d\zeta  -1)\sin(kd) + 2kd\cos(kd)]}{
4\zeta^{2}\sin(2 kd)+2\zeta k\sin(2kd)+k^{2}}\Big )\ .
\end{align}
These expressions are useful for examining various limits of the more general result. For instance, for  $q=0$ they yield $\mathcal{R}=\pi/e^2$,  in agreement with the result of B\"{u}ttiker. \cite{BUTTIKER2}

\section{Long cylinders}

\label{App1a}

When the cylinder is long  such that the frequency exceeds the level spacing,  $md^{2}\gg\omega^{-1}$,  a straightforward expansion of the ac conductance is not possible since the integrand in Eq. (\ref{acg}) is rapidly oscillating. For simplicity, we consider the case where the spin-orbit interact vanishes. Then the ac conductance, at zero temperature,  is
\begin{widetext}
\begin{align}
\label{gn}
{\cal G}(\omega )=\frac{e^{2}}{\pi\omega}\int_{E^{}_{F}-\omega}^{E^{}_{F}}dE&\sum_{q}\Big (1
- \frac{\zeta  \left(1-e^{2 i d k}\right)-i k }{\zeta  \left(1-e^{-2 i d k}\right)+i k }
\times
\frac{\zeta  \left(1-e^{-2 i d k^{}_{\omega}}\right)+i k^{}_{\omega} }{\zeta  \left(1-e^{2 i d k^{}_{\omega}}\right)-i k^{}_{\omega} }\Big )\ ,
\end{align}
with $k_{\omega}\equiv\sqrt{2m(E+\omega )-q^{2}}$.
The averaging over the rapid oscillations is carried out by integrating in the complex plane over the contour of the unit circle, 
\begin{align}
\label{gn1}
\frac{\pi\omega}{e^2}\mathcal{G}(\omega)&=\oint dz^{}_1 \oint d^{}z_2\int_{E^{}_{F}-\omega}^{E^{}_{F}}dE\sum_{q}
\Big (1- \frac{\zeta  \left(1-e^{z^{}_1}\right)-i k }{\zeta  \left(1-e^{-z^{}_1}\right)+i k }
 \times\frac{\zeta  \left(1-e^{-z^{}_2}\right)+i k_{\omega} }{\zeta  \left(1-e^{z^{}_2}\right)-i k^{}_{\omega} } \Big )\nonumber\\
&  = 
  \int_{E^{}_{F}-\omega}^{E^{}_{F}}dE\sum_{q} \left(1-\frac{(\zeta  -i k) }{(\zeta  +i k) } \frac{(\zeta  +i k^{}_{\omega}) }{(\zeta  -i k^{}_{\omega} )} \right)\ .
\end{align}
\end{widetext}
The low-frequency expansion of the ac conductance now yields
\begin{align}
{\cal C}=\frac{e^{2}}{2\pi}\sum_{q} \frac{4m\zeta}{k^{}_{\mu}(k^{2}_{\mu}+\zeta^2)}\ ,
\end{align}
and
\begin{align}
   \mathcal{R}=\frac{\pi}{e^2}\frac{\sum_q \left( \frac{4m\zeta}{k^{}_{\mu}(k_{\mu}^2+\zeta^2)}\right)^2}{\left(\sum_q \frac{4m\zeta}{k^{}_{\mu}(k_{\mu}^2+\zeta^2)}\right)^2}\ ,
\end{align}
where $k_{\mu}=\sqrt{2mE_{F}-q^{2}}$ (recall that at zero temperature $\mu=E_{F}$).
For the lowest transverse mode $q=0$ we find $\mathcal{R}=\pi/e^2$,  in agreement with the result of Refs. \onlinecite{Mora}. 
As noted  after Eq. (\ref{FF}), the effect of the spin-orbit interaction  can be gauged out for a single-channel scatterer
and therefore the result
 $\mathcal{R}=\pi/e^2$ holds also when the scattered electrons undergo a spin-orbit interaction.

\section{The scattering matrix of a tube}
\label{App2}

In order to derive the scattering matrix for the  geometry  depicted in Fig. \ref{figfull} we need to consider  the wave functions in the scattering region $|x|<d$ and in the two leads, the regions $x\geq d$ and $x\leq -d$. 
In the region $|x|<d$ we choose the same one as in Appendix \ref{App1}, Eqs. (\ref{phit}), (\ref{V}), and (\ref{v}). For the wave functions in the leads we write the Ansatz 
\begin{widetext}
\begin{align}
    \varphi^{}_L(x,y)&=\sum_q e^{iqy}\Big[ e^{i k (x+d)}|c^{}_{L,\rm{in}}\rangle
  +e^{-i k (x+d)}|c^{}_{L,\rm{out}}\rangle\Big] \ , \ \ x<-d\ ,\nonumber\\
  \varphi^{}_R(x,y)&=\sum_q e^{iqy}\Big[ e^{-i k (x-d)}|c^{}_{R,\rm{in}}\rangle
  +e^{i k (x-d)}|c^{}_{R,\rm{out}}\rangle\Big] \ ,\ \ x>d\ ,
\end{align}
where $|c_{\gamma,\rm{in}}\rangle$ ($|c_{\gamma,\rm{out}}\rangle$) is the incoming (outgoing) spinor in lead $\gamma$. Note that the direction of the incoming spinors is toward the central region, i.e., the scatterer.

As before, the transverse channels are not coupled and therefore we may solve the scattering matrix for a certain $q$.
We use the boundary conditions to eliminate the spinors in the scattering region $|x|\leq d$. The first set of boundary conditions follows from the continuity of the wave functions at $x=d$,
\begin{align}
\label{eq11}
    \mathbf{v^{}_u}e^{ik^{}_u d  \mathbf{\sigma^{}_z}}|c^{}_{\rm{u}}\rangle
  +\mathbf{v^{}_d}e^{ik^{}_d d\mathbf{\sigma^{}_z}}|c^{}_{\rm{d}}\rangle=|c^{}_{R,\rm{in}}\rangle
  +|c^{}_{R,\rm{out}}\rangle\ ,
\end{align}
and the continuity of the wave functions at $x=-d$, 
\begin{align}
   \mathbf{v^{}_u}e^{-ik^{}_u d  \mathbf{\sigma^{}_z}}|c^{}_{\rm{u}}\rangle
  +\mathbf{v^{}_d}e^{-ik^{}_d d\mathbf{\sigma^{}_z}}|c^{}_{\rm{d}}\rangle=
   |c^{}_{L,\rm{in}}\rangle
  +|c^{}_{L,\rm{out}}\rangle\ ,
\end{align}
where $k_{\rm u,d}$ is given in Eq. (\ref{kud}).  The second set of boundary conditions comes from  the continuity of the current, at $x=d$
\begin{align}
  &ik(|c^{}_{L,\rm{in}}\rangle-|c^{}_{L,\rm{out}}\rangle)+(2\zeta+i\alpha \mathbf{\sigma^{}_y})(|c^{}_{L,\rm{in}}\rangle+|c^{}_{L,\rm{out}}\rangle)=ik^{}_u\mathbf{v^{}_u}\mathbf{\sigma^{}_z}e^{-ik^{}_u d  \mathbf{\sigma^{}_z}} |c^{}_{u}\rangle +ik^{}_d \mathbf{v^{}_d}\mathbf{\sigma^{}_z }e^{-ik^{}_d d  \mathbf{\sigma^{}_z}}|c^{}_{d}\rangle \ ,
  \label{bc2r}
\end{align}
and at $x=-d$ 
\begin{align}
  &ik(|c^{}_{R,\rm{in}}\rangle-|c^{}_{R,\rm{out}}\rangle)+(2\zeta-i\alpha \mathbf{\sigma^{}_y})(|c^{}_{R,\rm{in}}\rangle+|c^{}_{R,\rm{out}}\rangle)=-ik^{}_u\mathbf{v^{}_u}\mathbf{\sigma_z}e^{ik^{}_u d  \mathbf{\sigma^{}_z}} |c^{}_{u}\rangle -ik^{}_d \mathbf{v^{}_d}\mathbf{\sigma^{}_z }e^{ik^{}_d d  \mathbf{\sigma^{}_z}}|c^{}_{d}\rangle\ .
  \label{bc2l}
\end{align}
Once the spinors $|c_{\rm u}\rangle$ and $|c_{\rm d}\rangle $ are eliminated, we find
\begin{align}
\left [\begin{array}{c}|c^{}_{L,{\rm out}}\rangle  \\   |c^{}_{R,{\rm out}}\rangle\end{array}\right ]={\cal S}
\left [\begin{array}{c}|c^{}_{L,{\rm in}} \rangle\\   |c^{}_{R,{\rm in}}\rangle\end{array}\right ]\ ,\ \ \ {\cal S}=\left [\begin{array}{cc}{\bf R}^{}_{LL}&{\bf T}^{}_{RL} \\{\bf T}^{}_{LR}&{\bf R}^{}_{RR}\end{array}\right]\ ,
\label{Sfull}
\end{align}
where
\begin{align}
    \mathbf{R}^{}_{LL} &= -1+2ik\left(\tilde{\mathbf{G}}-\tilde{\mathbf{F}}\mathbf{G}^{-1}\mathbf{F}\right)^{-1}\tilde{\mathbf{F}}\mathbf{G}^{-1} \ ,\ \ \   \mathbf{T}_{RL}=2ik\left(\tilde{\mathbf{G}}-\tilde{\mathbf{F}}\mathbf{G}^{-1}\mathbf{F}\right)^{-1}    \nonumber\\
    \mathbf{R}^{}_{RR} &= -1+2ik\left(\mathbf{G}-\mathbf{F}\tilde{\mathbf{G}}^{-1}\tilde{\mathbf{F}}\right)^{-1}\mathbf{F}\tilde{\mathbf{G}}^{-1} \ ,\ \ \ 
     \mathbf{T}_{LR}=2ik\left(\mathbf{G}-\mathbf{F}\tilde{\mathbf{G}}^{-1}\tilde{\mathbf{F}}\right)^{-1}\ .
\label{RT}
\end{align}
Here we have introduced the definitions
\begin{align}
    \mathbf{F}&=2\zeta^{}_L+i\alpha\sigma^{}_y-ik-ik^{}_{\rm u}\mathbf{v}^{}_{\rm u}\sigma^{}_ze^{-ik^{}_{\rm u}\sigma^{}_z d}\mathbf{B}_{\rm u}^{-1}-ik_{\rm d}^{}\mathbf{v}^{}_{\rm d} \sigma^{}_ze^{-ik^{}_{\rm d}\sigma^{}_z d}\mathbf{B}_{\rm d}^{-1} \ ,\nonumber\\
   \tilde{ \mathbf{F}}&=2\zeta^{}_R-i\alpha\sigma^{}_y-ik-ik^{}_{\rm u}\mathbf{v}^{}_{\rm u}\sigma^{}_ze^{ik^{}_{\rm u}\sigma^{}_z d}\mathbf{B}_{\rm u}^{-1}\mathbf{A}^{}_{\rm d}-ik_{\rm d}\mathbf{v}^{}_{\rm d} \sigma^{}_ze^{ik^{}_{\rm d}\sigma^{}_z d}\mathbf{B}_{\rm d}^{-1}\mathbf{A}^{}_{\rm u} \ ,\nonumber\\
   \mathbf{G}&=-ik^{}_{\rm u}\mathbf{v}^{}_{\rm u}\sigma^{}_ze^{-ik^{}_{\rm u}\sigma^{}_z d}\mathbf{B}_{\rm u}^{-1}\mathbf{A}^{}_{\rm d}-ik^{}_{\rm d}\mathbf{v}^{}_{\rm d}\sigma^{}_ze^{-ik^{}_{\rm d}\sigma^{}_z d}\mathbf{B}_{\rm d}^{-1}\mathbf{A}^{}_{\rm u}\ ,\nonumber \\
   \tilde{\mathbf{G}}&=-ik^{}_{\rm u}\mathbf{v}^{}_{\rm u}\sigma^{}_ze^{ik^{}_{\rm u}\sigma^{}_z d}\mathbf{B}_{\rm u}^{-1}-ik_{\rm d}\mathbf{v}_{\rm d}\sigma^{}_ze^{ik_{\rm d}\sigma^{}_z d}\mathbf{B}_{\rm d}^{-1}\ ,
\end{align}
with
\begin{align}
    \mathbf{A}_{\rm l}^{}&=\mathbf{v}^{}_{\rm l }e^{-2ik^{}_{\rm l}\sigma^{}_z d}\mathbf{v}_{\rm l}^{-1}\ ,\ \ \ \ \   
    \mathbf{B}^{}_{\bf l}=\mathbf{v}^{}_{\rm l} e^{-ik^{}_{\rm l}\sigma^{}_z d}-\mathbf{A}^{}_{\rm l'}\mathbf{v}^{}_{\rm l} e^{ik^{}_{\rm l}\sigma^{}_z d}\ ,
\end{align}
where l=u or d, and l$\neq$l'.

\end{widetext}

\end{document}